\documentclass[groupedaddress,a4paper,nofootinbib,tightenlines]{revtex4}
\usepackage{amsmath,amsthm,amssymb,float}
\usepackage[paper=a4paper,margin=3cm]{geometry}
\usepackage{graphicx}
\newcommand{\al}{\alpha}
\newcommand{\be}{\beta}
\newcommand{\de}{\delta}

\newcommand{\vep}{\varepsilon}
\newcommand{\ga}{\gamma}
\newcommand{\ka}{\kappa}
\newcommand{\la}{\lambda}

\newcommand{\si}{\sigma}

\newcommand{\vp}{\varphi}
\newcommand{\ze}{\zeta}
\newcommand{\La}{\Lambda}
\newcommand{\Si}{\Sigma}
\newcommand{\boe}{\mathbf{e}}
\newcommand{\bh}{\mathbf{h}}
\newcommand{\bk}{\mathbf{k}}

\newcommand{\bn}{\mathbf{n}}
\newcommand{\bp}{\mathbf{p}}
\newcommand{\bs}{\mathbf{s}}
\newcommand{\bx}{\mathbf{x}}

\newcommand{\bde}{{\boldsymbol{\de}}}

\newcommand{\bxi}{{\boldsymbol{\xi}}}

\newcommand{\bz}{\mathbf{z}}

\newcommand{\tH}{\tilde{H}}
\newcommand{\tHsc}{{\tilde{H}}^{\mathrm{sc}}}

\newcommand{\hH}{\hat{H}}

\newcommand{\hJ}{\widehat{J}}
\newcommand{\NN}{{\mathbb N}}
\newcommand{\RR}{{\mathbb R}}
\newcommand{\CC}{{\mathbb C}}

\newcommand{\cB}{{\mathcal B}}

\newcommand{\cE}{{\mathcal E}}
\newcommand{\cF}{{\mathcal F}}
\newcommand{\cH}{{\mathcal H}}

\newcommand{\cP}{{\mathcal P}}
\newcommand{\cR}{{\mathcal R}}

\newcommand{\cY}{{\mathcal Y}}

\newcommand{\cZ}{{\mathcal Z}}
\def\Esc{E^{\mathrm{sc}}}
\def\Hsc{H^{\mathrm{sc}}}
\def\Lasc{\La^{\mathrm{sc}}}
\def\Zsc{Z^{\mathrm{sc}}}
\newcommand{\pa}{\partial}

\newcommand{\id}{1\hspace{-.25em}{\rm I}}

\def\ket#1{|#1\rangle}

\newcommand{\ms}{\mspace{1mu}}
\renewcommand{\le}{\leqslant}
\renewcommand{\ge}{\geqslant}
\renewcommand{\leq}{\leqslant}
\renewcommand{\geq}{\geqslant}

\newcommand{\cEmin}{\cE_{\mathrm{min}}}
\newcommand{\cEmax}{\cE_{\mathrm{max}}}
\newcommand{\erf}{\operatorname{erf}}

\newcommand{\tr}{\operatorname{tr}}

\newcommand{\card}{\operatorname{card}}

\newcommand{\e}{\mathrm{e}}

\newcommand{\diff}{\mathrm{d}}
\newcommand{\smax}{s_{\mathrm{max}}}

\newcounter{ex}
\def\cond{\stepcounter{ex}\hskip-.75cm
\makebox[.55cm][r]{(\roman{ex})}\hskip.2cm}
\begin{document}
\title{Inozemtsev's hyperbolic spin model and its related spin chain}
\author{J.C. \surname{Barba}}%
\author{F. \surname{Finkel}}%
\author{A. \surname{Gonz\'alez-L\'opez}}\email{artemio@fis.ucm.es}%
\author{M.A. \surname{Rodr\'\i guez}}%
\affiliation{Departamento de F\'\i sica Te\'orica II, Universidad Complutense, 28040 Madrid,
  Spain}
  \date{3 May 2010}
\begin{abstract}
  In this paper we study Inozemtsev's $\mathrm{su}(m)$ quantum spin model with hyperbolic
  interactions and the associated spin chain of Haldane--Shastry type introduced by Frahm and
  Inozemtsev. We compute the spectrum of Inozemtsev's model, and use this result and the freezing
  trick to derive a simple analytic expression for the partition function of the Frahm--Inozemtsev
  chain. We show that the energy levels of the latter chain can be written in terms of the usual
  motifs for the Haldane--Shastry chain, although with a different dispersion relation. The
  formula for the partition function is used to analyze the behavior of the level density and the
  distribution of spacings between consecutive unfolded levels. We discuss the relevance of our
  results in connection with two well-known conjectures in quantum chaos.
\end{abstract}

\maketitle


\section{Introduction}\label{sec.intro}

Over the last few years a significant amount of effort has been devoted to the study of spin
chains of Haldane--Shastry (HS) type, due to their remarkable integrability properties and their
interest in connection with several important conjectures in quantum chaos. This class of chains,
intimately related to integrable dynamical models of Calogero--Sutherland
type~\cite{Ca71,Su71,Su72}, are characterized by the fact that the interactions between the spins
are both long-ranged and position-dependent. For instance, in the original Haldane--Shastry
chain~\cite{Ha88,Sh88} the spins occupy equidistant positions on a circle, the strength of the
interactions being inversely proportional to the square of the distance between the spins measured
along the chord. Historically, this model was introduced while searching for a spin chain whose
exact ground state coincided with Gutzwiller's variational wavefunction for the one-dimensional
Hubbard model in the limit of large on-site interaction~\cite{Hu63,Gu63,GV87}. In fact, the HS
chain can be obtained in this limit from the Hubbard model with long-range hopping in the
half-filling regime~\cite{GR92}. In Haldane's original paper~\cite{Ha88}, the spectrum of the HS
chain with spin $1/2$ was inferred on the basis of numerical calculations. In particular, it was
observed that the levels are highly degenerate, which suggests the presence of a large underlying
symmetry group and the possible integrability of the model. This symmetry group was subsequently
identified~\cite{HHTBP92} as the Yangian $\cY(\mathrm{sl}_m)$, where $m$ is the number of internal
degrees of freedom. As to the model's integrability, it was established around the same time by
Fowler and Minahan~\cite{FM93} using Polychronakos's exchange operator formalism~\cite{Po92}.

The rigorous derivation of the spectrum of the HS chain with arbitrary $\mathrm{su}(m)$ spin was
carried out by Bernard et al.~\cite{BGHP93} by taking advantage of its connection with the
generalization of Sutherland's model to particles with spin~\cite{HH92}. At the heart of this
connection is the mechanism known as Polychronakos's ``freezing trick''~\cite{Po93}. The physical
idea behind this mechanism is that when the coupling constant of the spin Sutherland
(trigonometric) model tends to infinity, the particles concentrate around the equilibrium of the
scalar part of the potential, so that the dynamical and internal degrees of freedom decouple. It
can be shown that the coordinates of this equilibrium are essentially the HS chain sites, and that
in this limit the internal degrees of freedom are governed by the chain's Hamiltonian. The
freezing trick can also be applied to the spin Calogero (rational) model~\cite{MP93}, obtaining in
this way a spin chain ---the so-called Polychronakos--Frahm (PF) chain~\cite{Po93,Fr93}--- with
non-equidistant sites given by the zeros of the $N$-th degree Hermite polynomial, $N$ being the
number of spins.

The original Calogero and Sutherland models mentioned above are both based on the root system
$A_{N-1}$, in the sense that the interaction between the particles depends only on their relative
distance. As shown by Olshanetsky and Perelomov~\cite{OP83}, there are integrable generalizations
of these models associated with any (extended) root system, the rank of the root system basically
coinciding with the number of particles. For this reason, the most studied models of
Calogero--Sutherland (CS) type are by far those associated with the $BC_N$ root system (including
the hyperbolic Sutherland model)~\cite{OP83,Ya95,FGGRZ01,FGGRZ03,EFGR05}, and to a lesser extent
with the $D_N$ system~\cite{BFG09,BFG09pre}. A hyperbolic variant of the Sutherland model of
$A_{N-1}$ type with an external confining potential of Morse type has also been considered in the
literature, in both the scalar~\cite{IM86} and the spin~\cite{In96} cases.

For all the spin CS models mentioned in the previous paragraphs, a corresponding spin chain of
Haldane--Shastry type has been constructed by means of the freezing
trick~\cite{FI94,Ya95,YT96,FGGRZ03,BFG09,BFG09pre}. In the case of the rational and trigonometric
chains, this mechanism has been applied to derive a closed-form expression for the partition
function in terms of the quotient of the partition functions of the corresponding spin and scalar
dynamical models~\cite{Po94,EFGR05,FG05,BFGR08,BFG09,BFG09pre}, whose spectrum can be easily
computed. Expanding the partition function in powers of $q\equiv\e^{-1/(k_{\mathrm B}T)}$, one can
compute the chain's spectrum for relatively large values of $N$ and determine some of its
statistical properties. A common feature of all of these chains is the fact that when the number
of sites is sufficiently large the level density is approximately Gaussian. This result, for which
there is ample numerical evidence, has also been rigorously established in some
cases~\cite{EFG09}. The knowledge of a continuous approximation to the (cumulative) level density
is of great importance in the context of quantum chaos, as it is used to transform the raw
energies so that the resulting ``unfolded'' spectrum has an approximately uniform level
density~\cite{GMW98}. The distribution of spacings between consecutive unfolded levels is widely
used for testing the integrable vs.~chaotic character of a quantum system. Indeed, according to a
long-standing conjecture of Berry and Tabor~\cite{BT77}, the spacings distribution $p(s)$ of a
``generic'' quantum system whose classical counterpart is integrable should follow Poisson's law
$p(s)=\e^{-s}$. On the other hand, the Bohigas--Giannoni--Schmidt conjecture~\cite{BGS84} asserts
that the spacings distribution of a fully chaotic quantum system is given by Wigner's surmise
$p(s)=(\pi s/2)\exp(-\pi s^2/4)$, characteristic of the Gaussian orthogonal ensemble (GOE) in
random matrix theory. Both of these conjectures have been shown to hold in many different systems,
both in the integrable~\cite{PZBMM93,AMV02} and fully chaotic cases~\cite{GMW98}. Rather
surprisingly, the spacings distribution of all the integrable spin chains of HS type studied so
far is neither of Poisson's nor Wigner's type~\cite{FG05,BB06,BFGR08,BFGR08epl,BFGR09,BFG09}.
Thus, at least in this respect, spin chains of HS type appear to be exceptional among the class of
integrable systems.

Unlike their rational or trigonometric counterparts, HS chains with hyperbolic interactions have
received comparatively less attention in the literature. In particular, the spectrum of the
hyperbolic $BC_N$ chain is not known, whereas that of the $A_{N-1}$ chain has been
conjectured~\cite{FI94} only for spin $1/2$. The main purpose of this paper is precisely to fill
this gap for the hyperbolic chain of $A_{N-1}$ type, which we shall henceforth refer to as the
Frahm--Inozemtsev (FI) chain. Although at the formal level this chain is the hyperbolic analog of
the original Haldane--Shastry chain, in practice both chains turn out to be quite different.
Indeed, while the sites of the HS chain are equidistant on a circle, those of the FI chain are not
and, moreover, depend on a parameter $\be$ in a nontrivial way. We shall see that the energies of
the FI chain also depend on this parameter, which complicates the statistical analysis of the
spectrum. An important consequence of this dependence is the fact that for generic values of $\be$
the spacings distribution is qualitatively different from that of the integrable spin chains of HS
type studied so far. At any rate, for all values of $\be$ the spacings distribution of the FI
chain is neither of Poisson's nor Wigner's type, in spite of the fact that this chain is probably
integrable~\cite{FI94}.

The paper is organized as follows. In Section~\ref{sec.themodels} we recall the definition of
Inozemtsev's hyperbolic spin dynamical model and its scalar counterpart, and explicitly construct
the Frahm--Inozemtsev spin chain by applying the freezing trick to these models.
Section~\ref{sec.specdyn} is devoted to the computation of the spectrum of Inozemtsev's spin
dynamical model, which was partially known only in the case of spin $1/2$. Our approach is based
on relating the Hamiltonian to an auxiliary differential-difference operator, which we
triangularize by expressing it in terms of suitable Dunkl--Cherednik operators of type
A~\cite{Du89,Ch94}. In Section~\ref{sec.partfunc} we use the freezing trick and the results of the
previous section to derive a closed-form expression for the partition function of the
$\mathrm{su}(m)$ FI chain. Using this expression and some general results for other spin chains of
$A_{N-1}$ type~\cite{BBHS07,BBS08}, we obtain a simple formula for the spectrum in terms of the
usual \emph{motifs}~\cite{HHTBP92}. In particular, this provides a rigorous proof of Frahm and
Inozemtsev's conjecture for the spectrum in the spin $1/2$ case. With the help of the partition
function, in Section~\ref{sec.spectrum} we analyze several statistical properties of the chain's
spectrum. When $\be=O(N)$, our numerical computations show that the level density is Gaussian as
the number of sites tends to infinity. Taking as the spectrum unfolding function the cumulative
Gaussian distribution, we have also studied the density of spacings for large $N$ and different
values of the spin when the parameter $\be$ is $O(N)$. Our calculations clearly indicate that the
density of spacings exhibits the behavior previously found in other chains of HS type only when
$\be$ is an integer or a rational with a ``small'' denominator. The paper ends with two technical
Appendices in which we prove the existence of a unique solution of the system defining the chain
sites, and derive a closed-form expression for the mean and variance of the chain's energies.

\section{The models}\label{sec.themodels}

In this section we describe Inozemtsev's hyperbolic spin dynamical model~\cite{In96} and its
corresponding spin chain of Haldane--Shastry type~\cite{FI94}, whose study is the purpose of this
paper. The Hamiltonian of the spin dynamical model is given by
\begin{equation}
  H = -\triangle + b^2\sum_i(\e^{-2x_i}-1)^2+a\sum_{i\neq j}\frac{a-\vep S_{ij}}{\sinh^2(x_i-x_j)}\,,
  \label{H}
\end{equation}
where the sums run from $1$ to $N$ (as always hereafter, unless otherwise stated), $a>1/2$, $b>0$,
$\vep=\pm1$, and the operators $S_{ij}$ permute the spins of the $i$-th and $j$-th particles. More
precisely, let $\Si=(\CC^m)^{\otimes N}$ be the space of internal degrees of freedom, and denote
by $\ket\bs\equiv\ket{s_1,\dots,s_N}$, with $s_i\in\{1,\dots,m\}$, an element of the canonical
basis of $\Si$. The action of $S_{ij}$ on $\ket\bs$ is then given by
\[
S_{ij}\ket{\dots,s_i,\dots,s_j,\dots}=\ket{\dots,s_j,\dots,s_i,\dots}\,.
\]
It is well known that the operators $S_{ij}$ can be expressed in terms of the generators $t^\al_k$
of the fundamental representation of $\mathrm{su}(m)$ for the $k$-th particle as
\[
S_{ij}=2\sum_{\al=1}^{m^2-1}t_i^\al t_j^\al+\frac1m\,,
\]
where we have used the normalization $\tr(t^\al_kt^{\ga}_k)=\frac12\de^{\al\ga}$. It was shown in
Ref.~\cite{In96} that the above model is completely integrable for arbitrary $m$. Regarding its
spectrum, in the latter reference only a few eigenstates (including the ground state) and their
corresponding energies were computed in the special case of spin $1/2$ ($m=2$). The
Hamiltonian~\eqref{H} is the spin version of the scalar model
\begin{equation}
  \Hsc = -\triangle + b^2\sum_i(\e^{-2x_i}-1)^2+a(a-1)\sum_{i\neq j}\sinh^{-2}(x_i-x_j)\,,
  \label{Hsc}
\end{equation}
previously studied in Ref.~\cite{IM86}. In particular, in the latter reference the spectrum of
$\Hsc$ was computed in closed form, together with the eigenfunctions of the ground state and
several excited states. It should be noted that, due to the impenetrable nature of the
singularities of $H$ and $\Hsc$ in the hyperplanes $x_i=x_j$ ($i\ne j$), the configuration space
of both Hamiltonians~\eqref{H} and~\eqref{Hsc} should be taken as one of the Weyl chambers of
$A_{N-1}$ type, say
\begin{equation}\label{C}
  C=\{\bx\equiv(x_1,\dots,x_N)\in\RR^N\mid x_1<x_2<\dots<x_N\}\,.
\end{equation}

We shall next apply Polychronakos's ``freezing trick''~\cite{Po93} to the spin dynamical
model~\eqref{H} in order to construct its associated spin chain of Haldane--Shastry type. To this
end, we rescale the strength of the Morse potential as
\[
b=\be a,
\]
and consider the strong coupling limit $a\to\infty$. The Hamiltonian~\eqref{H} can be written as
\[
H=-\triangle+a^2 U(\bx)+O(a)\,,
\]
where the scalar potential $U$ is given by
\begin{equation}\label{U}
  U(\bx)=\be^2\sum_i(\e^{-2x_i}-1)^2+\sum_{i\neq j}\sinh^{-2}(x_i-x_j)\,.
\end{equation}
We shall show in Appendix~\ref{sec.minimum} that $U$ possesses a minimum in the set $C$ if and
only if
\begin{equation}\label{cond}
  \be>2(N-1)\,,
\end{equation}
and that this minimum is in fact unique. It follows that, as $a\to\infty$, the eigenfunctions of
$H$ become sharply peaked around the unique minimum $\bxi\equiv(\xi_1,\dots,\xi_N)$ of $U$ in $C$.
Since
\begin{equation}\label{H-chain}
  H=\Hsc+4a\,h(\bx)\,,
\end{equation}
where
\begin{equation}
  h(\bx)=\frac14\,\sum_{i\neq j}\frac{1-\vep S_{ij}}{\sinh^2(x_i-x_j)}\,,
  \label{h}
\end{equation}
in the limit $a\to\infty$ the dynamical and internal degrees of freedom decouple, the latter being
governed by the Hamiltonian
\begin{equation}
  \cH\equiv h(\bxi)=\frac14\,\sum_{i\neq j}\frac{1-\vep S_{ij}}{\sinh^2(\xi_i-\xi_j)}\,.
  \label{cH}
\end{equation}
As shown in Appendix~\ref{sec.minimum}, the chain sites $\xi_i$ can be expressed in terms of the
zeros $\ze_1<\cdots<\ze_N$ of the generalized Laguerre polynomial $L_N^{\be-2N+1}$ as
\begin{equation}
  \label{zei}
  2\xi_i=\log\be-\log\ze_{N-i+1}\,.
\end{equation}
Recall, in this respect, that Eq.~\eqref{cond} is precisely the condition which guarantees that
the zeros of $L_N^{\be-2N+1}$ are positive and distinct~\cite{Sz75}. In fact, since $\cH$ is
invariant under $\xi_i\mapsto-\xi_i+c$, with $c$ a constant, we can alternatively define the sites
of the chain~\eqref{cH} by the simpler formula
\begin{equation}\label{cHsites}
  \xi_i=\frac 12\log\ze_i\,.
\end{equation}
We can also express the Hamiltonian~\eqref{cH} directly in terms of the zeros $\ze_i$ of
$L_N^{\be-2N+1}$ as
\begin{equation}
  \label{cHz}
  \cH = \sum_{i\neq j}\frac{\ze_i\ze_j}{(\ze_i-\ze_j)^2}\,(1-\vep S_{ij})\,.
\end{equation}
We shall take Eqs.~\eqref{cH}-\eqref{cHsites} as the definition of the Hamiltonian of the
Frahm--Inozemtsev chain, although it differs from the normalization used in Ref.~\cite{FI94} by
a factor of~$2$. If $\vep$ takes the value $1$ (respectively $-1$) the corresponding FI chain is
of ferromagnetic (respectively antiferromagnetic) type. We shall sometimes use the more precise
notation $\cH^{\mathrm F}$ (respectively $\cH^{\mathrm{AF}}$) to denote the Hamiltonian of the
ferromagnetic (respectively antiferromagnetic) chain. Note finally that, unlike other spin chains
of HS type associated with the $A_{N-1}$ root system, the Hamiltonian of the
FI chain depends on an essential parameter $\be>2(N-1)$ through the zeros $\ze_i$.

\section{Spectrum of the dynamical models}\label{sec.specdyn}

In this section we shall compute the point spectrum of Inozemtsev's spin dynamical model~\eqref{H}
and of its scalar counterpart~\eqref{Hsc}. As is customary when studying quantum
Calogero--Sutherland models with spin, we introduce the auxiliary \emph{scalar} operator
\begin{equation}
  \label{Hp}
  \hH = -\triangle+b^2\sum_i(\e^{-2x_i}-1)^2+a\sum_{i\neq j}\frac{a-K_{ij}}{\sinh^2(x_i-x_j)}\,,
\end{equation}
where $K_{ij}$ is the coordinate permutation operator defined by
\begin{equation}
  (K_{ij}f)(x_1,\dots,x_i,\dots,x_j,\dots,x_N)=f(x_1,\dots,x_j,\dots,x_i,\dots,x_N)\,.
\end{equation}
The operator $\hH$ is naturally defined on (a suitable dense subset of) the Hilbert space
$L^2(\RR^N)$. On the other hand, due to the nature of their singularities on the hyperplanes
$x_i=x_j$ (with $i\ne j$), the Hamiltonians $H$ and $\Hsc$ act on (appropriate dense subsets of)
the Hilbert spaces $L^2(C)\otimes\Si$ and $L^2(C)$, respectively. However, proceeding as in
Ref.~\cite{BFG09pre}, one can show that these Hamiltonians are isospectral to their extensions
$\tH$ and $\tHsc$ to $\La\big(L^2(\RR^N)\otimes\Si\big)$ and $\Lasc\big(L^2(\RR^N)\big)$,
respectively, where $\La$ and $\Lasc$ denote the projectors onto spin and scalar states with any
fixed symmetry under particle permutations. We shall choose these extensions so that
\begin{equation}\label{tHs}
\tH=\hH\otimes\id\,\big|_{\La(L^2(\RR^N)\otimes\Si)}\,,\qquad
\tHsc=\hH\,\big|_{\Lasc(L^2(\RR^N))}\,.
\end{equation}
This is equivalent to the requirement that
\[
K_{ij}\La = \vep S_{ij}\La\,,\qquad K_{ij}\Lasc = \Lasc\,,
\]
so that $\La$ must project onto spin states with parity $\vep$ under particle permutations, while
$\Lasc$ is the symmetrizer under coordinate permutations.

\subsection{Dunkl operators}

In view of the above remarks, in order to compute the point spectra of $\tH$ and $\tHsc$ it is
enough to solve the analogous problem for the auxiliary operator $\hH$ in $L^2(\RR^N)$. To this
end, we introduce the gauged Dunkl operators
\begin{subequations}\label{Dunkl}
  \begin{align}
    \label{Dunklm}
    2\hJ^-_i &= e^{-2x_i}\Big[\partial_{x_i}-b\e^{-2x_i} -a\sum_{j\neq
      i}\big(1+\coth(x_i-x_j)\big)K_{ij} +b-1\Big]\,,\\
    \label{Dunkl0}
    2\hJ^{\,0}_i &= \partial_{x_i}-b\e^{-2x_i}-a\sum_{j\neq
      i}\big(1+\coth(x_i-x_j)\big)K_{ij}+2a\sum_{j<i}K_{ij}\,,
  \end{align}
\end{subequations}
in terms of which
\begin{equation}
  \label{Hhat}
  \hH =  -4\sum_i\big(\hJ_i^{\,0}\big)^2-4b\sum_i\,\hJ_i^-+N b^2\,.
\end{equation}
The operators~\eqref{Dunkl} are related to the type A Dunkl--Cherednik operators \cite{Du89,Ch94}
\begin{subequations}\label{Dunklz}
  \begin{align}
    J_i^{\ms-}&=\partial_{z_i}+a\sum_{j\neq i}\frac{1}{z_i-z_j}\,(1-K_{ij})\,,\label{Dunklmz}\\
    J_i^{\,0}&=z_i\,\partial_{z_i}+a\sum_{j\neq i}\frac{z_j}{z_i-z_j}\,(1-K_{ij})
    -a\sum_{j>i}K_{ij}+a(N-1)+\frac{1-b}2\label{Dunkl0z}
  \end{align}
\end{subequations}
by the gauge transformation
\[
\hJ^{\,-}_i=\rho\,J_i^-\,\rho^{-1}\,,\qquad \hJ^{\,0}_i=\rho\,J_i^0\,\rho^{-1}\,,
\]
where $z_i=\e^{2x_i}$ and $\rho$ is the ground state of the scalar Hamiltonian $\Hsc$, given by
\begin{equation}\label{rho}
  \rho = \exp\Big[\big(1+a(N-1)-b\big)\sum_ix_i-\frac b2\sum_i\e^{-2x_i}\Big]\,\prod_{i<j}|\sinh(x_i-x_j)|^a\,.
\end{equation}
It is well-known~\cite{FGGRZ01} that the operators~\eqref{Dunklz} preserve the polynomial modules
\[
\cR_n = \Big\{\prod_i z_i^{n_i}\mid n_i\in\NN_0\,,\ n_i\leq n\Big\}
\]
for arbitrary $n\in\NN_0\equiv\NN\cup\{0\}$. Proceeding as in Ref.~\cite{IM86}, we seek the
eigenfunctions of $\hH$ in $\rho\ms\cR_n$, where $n$ must be chosen so that $\rho\ms\cR_n\subset
L^2(\RR^N)$. Taking into account that
\[
  \prod_{i\neq j}\sinh(x_i-x_j) \propto \prod_{i\neq j}\e^{x_i+x_j}\cdot \prod_{i\neq
    j}\big(\e^{-2x_j}-\e^{-2x_i}\big) = \prod_i\e^{2(N-1)x_i}\cdot \prod_{i\neq
    j}\big(\e^{-2x_j}-\e^{-2x_i}\big)
\]
we have
\[
\rho\prod_iz_i^{n_i} \propto\prod_i \exp\Big[\left(2n_i+2a(N-1)+1-b\right)x_i-\frac
b2\e^{-2x_i}\Big] \cdot\prod_{i< j}\big|\e^{-2x_i}-\e^{-2x_j}\big|^{a}\,,
\]
and thus $\rho\prod_iz_i^{n_i}$ is square-integrable provided that
\[
n_i<\frac12\,(b-1)-a(N-1)\,,\qquad 1\leq i\leq N\,.
\]
Hence the maximum value of $n$ such that $\rho\ms\cR_n\subset L^2(\RR^N)$ is given by
\begin{equation}
  \label{nmax}
  n= \max\bigg\{n'\in\NN_0\mid n'<\frac12\,(b-1)-a(N-1)\bigg\}\,.
\end{equation}
Note, in particular, that $\hH$ possesses bound states if and only if the strength of the Morse
potential satisfies the condition
\begin{equation}\label{bmin}
  b>2a(N-1)+1\,.
\end{equation}

\subsection{Triangularization of $\hH$}

In order to compute the point spectrum of the auxiliary operator $\hH$, we shall construct a
non-orthonormal basis of $\rho\ms\cR_n$ (with $n$ given by Eq.~\eqref{nmax}) on which this operator
acts triangularly. To this end, it suffices to construct a basis of $\cR_n$ on which the gauge
transformed operator
\begin{equation}
  H'\equiv\rho^{-1}\,\hH\,\rho=-4\sum_i\big(J_i^{\,0}\big)^2-4b\sum_i\,J_i^-+N b^2
  \label{HpJs}
\end{equation}
(cf.~Eq.~\eqref{Hhat}) is represented by a triangular matrix. In fact, the elements of this basis
are simply the monomials
\[
\phi_{\bn}=\prod_i z_i^{n_i}\,,\qquad\bn\equiv(n_1,\dots,n_N)\in\NN_0^N\,,
\]
ordered as we shall now explain. Given a multiindex $\bn\in\NN_0^N$, we define the associated
non-increasing multiindex
\[ [\bn]=(n_{i_1},\dots,n_{i_N})\,,\qquad \text{with}\enspace n_{i_1}\geq\cdots\geq n_{i_N}\,.
\]
For $\bp,\bp'\in[\NN_0^N]$, we shall write $\bp\prec\bp'$ if the first nonzero component of
$\bp'-\bp$ is positive. In the case of two arbitrary multiindices $\bn,\bn'\in \NN_0^N$, we define
\[
\bn\prec\bn'\quad\Longleftrightarrow\quad[\bn]\prec[\bn']\,.
\]
Thus, for instance,
\[
(1,2,3,2)\prec(1,2,2,6)\prec(1,1,6,3)\,,
\]
since clearly
\[
(3,2,2,1)\prec(6,2,2,1)\prec(6,3,1,1)\,.
\]
Finally, we shall set $\phi_{\bn}\prec\phi_{\bn'}$ if and only if $\bn\prec\bn'$. This defines a
\emph{partial} order $\prec$ in the set of all monomials $\phi_{\bn}\in\cR_n$. We shall see that
the auxiliary operator $H'$ is represented by an upper triangular matrix in the basis $\cB$ of
$\cR_n$ consisting of the monomials $\phi_{\bn}$, ordered in any way consistent with the relation
$\prec$. The proof of this fact can be divided into four steps.

First of all, it is easy to show that the operators $J_i^-$ are strictly upper triangular with
respect to the basis $\cB$, i.e.,
\begin{equation}\label{Jmin}
  J_i^-\,\phi_\bn = \sum_{\bn'\prec\bn} c_{i,\bn'\bn}^-\ms\phi_{\bn'}\,.
\end{equation}
Indeed,
\[
\phi^{-1}_\bn J_i^-\,\phi_\bn=\frac{n_i}{z_i}+a\sum_{j\neq i}
\frac{1}{z_i-z_j}\,(1-z_i^{n_j-n_i}z_j^{n_i-n_j})\,.
\]
Since the terms in the sum over $j\neq i$ vanish when $n_i=n_j$, the latter sum can be written as
\begin{multline*}
  \sum_{j;\,n_j<n_i}\frac1{z_i}\,\frac{(z_j/z_i)^{n_i-n_j}-1}{(z_j/z_i)-1}-
  \sum_{j;\,n_j>n_i}\frac1{z_j}\,\frac{(z_i/z_j)^{n_j-n_i}-1}{(z_i/z_j)-1}\\
  = \sum_{j;\,n_j<n_i}\sum_{k=0}^{n_i-n_j-1}z_j^kz_i^{-k-1}-
  \sum_{j;\,n_j>n_i}\sum_{k=0}^{n_j-n_i-1}z_i^kz_j^{-k-1}\,.
\end{multline*}
Thus
\[
J_i^-\phi_{\bn} = n_i\phi_{\bn-\boe_i}+a\sum_{j;\,n_j<n_i}\sum_{k=0}^{n_i-n_j-1}
\phi_{\bn-(k+1)\boe_i+k\boe_j}-a\sum_{j;\,n_j>n_i}\sum_{k=0}^{n_j-n_i-1}
\phi_{\bn+k\boe_i-(k+1)\boe_j},
\]
where $\boe_k$ is the $k$-th element of the canonical basis of $\RR^N$. It can be easily checked
that the vectors $\bn-\boe_i$, $\bn-(k+1)\boe_i+k\boe_j$ and $\bn+k\boe_i-(k+1)\boe_j$ appearing
in the previous expression all precede $\bn$, which establishes our claim.

Consider now a non-decreasing multiindex $\bp\in[\NN_0^N]$, and set
\[
\ell(p_i)=\min\big\{j\mid p_j=p_i\big\}\,,\qquad \#(p_i)=\card\big\{j\mid p_j=p_i\big\}\,.
\]
The second step in the proof of the triangular character of $H'$ with respect to the basis $\cB$
consists in showing that
\begin{equation}
  \label{Ji0bp}
  J_i^{\,0}\phi_{\bp} = \lambda_i(\bp)\ms\phi_{\bp}+\sum_{\bn'\prec\bp}c_{i,\bn'\bp}^0\ms\phi_{\bn'}\,,
\end{equation}
where
\begin{equation}
  \label{lai}
  \lambda_i(\bp) = p_i+\frac12(1-b)+a\big(N+i+1-\#(p_i)-2\ell(p_i)\big)\,.
\end{equation}
Indeed, proceeding as before we obtain:
\begin{align*}
  \frac1a\,&\phi_{\bp}^{-1}\Big[J_i^{\,0}-a(N-1)+\frac
  12\,(b-1)-p_i\Big]\phi_{\bp}\\
  &= -\sum_{j<i}\frac{(z_i/z_j)^{p_j-p_i}-1}{(z_i/z_j)-1}
  +\sum_{j>i}\frac{z_j}{z_i}\,\frac{(z_j/z_i)^{p_i-p_j}-1}{(z_j/z_i)-1}
  -\sum_{j>i}\bigg(\frac{z_j}{z_i}\bigg)^{p_i-p_j}\\
  &=\sum_{j>i;\,p_j<p_i}\sum_{k=1}^{p_i-p_j-1}\bigg(\frac{z_j}{z_i}\bigg)^k
  -\sum_{j<i;\,p_j>p_i}\sum_{k=0}^{p_j-p_i-1}\bigg(\frac{z_i}{z_j}\bigg)^k
  -\card\{j>i\mid p_j=p_i\}\\
  &= \sum_{j>i;\,p_j<p_i}\sum_{k=1}^{p_i-p_j-1}\bigg(\frac{z_j}{z_i}\bigg)^k
  -\sum_{j<i;\,p_j>p_i}\sum_{k=1}^{p_j-p_i-1}\bigg(\frac{z_i}{z_j}\bigg)^k\\[1mm]
  &\hphantom{\sum_{j>i;\,p_j<p_i}\sum_{k=1}^{p_i-p_j-1}\bigg(\frac{z_j}{z_i}\bigg)^k{}-{}}
  -\card\{j<i\mid p_j>p_i\}-\card\{j>i\mid p_j=p_i\}\,.
\end{align*}
Taking into account that
\[
\card\{j<i\mid p_j>p_i\}=\ell(p_i)-1,\quad\card\{j>i\mid
p_j=p_i\}=\#(p_i)+\ell(p_i)-i-1
\]
we have
\[
  J_i^{\,0}\phi_{\bp}=\lambda_i(\bp)\ms\phi_{\bp}+a\sum_{j>i;\,p_j<p_i}\sum_{k=1}^{p_i-p_j-1}
  \phi_{\bp-k\boe_i+k\boe_j}
  -a\sum_{j<i;\,p_j>p_i}\sum_{k=1}^{p_j-p_i-1}\phi_{\bp+k\boe_i-k\boe_j}\,,
\]
which obviously proves our claim.

Consider next the action of the operator $J_i^0$ on a basis function $\phi_{\bn}$ with arbitrary
$\bn\in\NN_0^N$. A computation totally analogous to the previous one shows that
\begin{multline*}
  \frac1a\,\Big[J_i^0-n_i-a(N-1)+\frac12(b-1)+a\card\{j\mid n_j>n_i\}+a\card\{j>i\mid
  n_j=n_i\}\Big]\phi_{\bn}\\
  =\sum_{j<i;\,n_j<n_i}\phi_{K_{ij}\bn}\,
  -\sum_{j>i;\,n_j>n_i}\phi_{K_{ij}\bn}
  +\sum_{j;\,n_j<n_i}\sum_{k=1}^{n_i-n_j-1}\phi_{\bn-k\boe_i+k\boe_j}
  -\sum_{j;\,n_j>n_i}\sum_{k=1}^{n_j-n_i-1}\phi_{\bn+k\boe_i-k\boe_j}\,.
\end{multline*}
The first two sums in the last expression involve basis functions with multiindices $\bn'$ such
that $[\bn']=[\bn]$, while all the multiindices appearing in the last two sums precede $\bn$.
Hence we have
\begin{equation}
  \label{Ji0bn}
  J_i^{\,0}\phi_{\bn} = \sum_{\bn';\,[\bn']\preceq[\bn]}\tilde c_{i,\bn'\bn}^{\,0}\phi_{\bn'}\,.
\end{equation}

Although the last identity indicates that the operators $J_i^0$ need not be triangular with
respect to the basis $\cB$, we shall next show that the sum $\sum_i(J_i^0)^2$ appearing in
$H'$ is upper triangular in the latter basis. This fact, together with Eq.~\eqref{Jmin}, implies
that $H'$ is represented by an upper triangular matrix in the basis $\cB$.

Indeed, given $\bn\in\NN_0^N$ let $P$ be any permutation such that $\bn=P[\bn]$, and (with a
slight abuse of notation) denote also by $P$ the linear operator defined by
$P\phi_{\bn'}=\phi_{P\bn'}$, for all $\bn'\in\NN_0^N$. Since
$(P\phi_{\bn'})(\bz)=\phi_{\bn'}(P^{-1}\bz)$, the operator $\sum_i(J_i^{\,0})^2$
obviously commutes with $P$, and therefore
\[
\sum_i\big(J_i^{\,0}\big)^2\phi_{\bn}=P\sum_i\big(J_i^{\,0}\big)^2\phi_{[\bn]}\,.
\]
Calling $[\bn]=\bp$ and using Eqs.~\eqref{Ji0bp} and~\eqref{Ji0bn} we easily obtain
\begin{align*}
  \sum_i\big(J_i^{\,0}\big)^2\phi_{\bp}&=\sum_iJ_i^0\Big(\la_i(\bp)\ms\phi_{\bp}
  +\sum_{\bn'\prec\bp}c_{i,\bn'\bp}^0\phi_{\bn'}\Big)\\
  &=\sum_i\la_i(\bp)^2\phi_{\bp}+\sum_i\sum_{\bn'\prec\bp}\la_i(\bp)\ms
  c_{i,\bn'\bp}^0\phi_{\bn'}
  +\sum_i\sum_{\substack{\bn'\prec\bp\\ [\bn'']\preceq[\bn']}}c_{i,\bn'\bp}^0\tilde
  c_{i,\bn''\bn'}^0\phi_{\bn''}\,,
\end{align*}
and hence
\[
\sum_i\big(J_i^{\,0}\big)^2\phi_{\bn}
=\sum_i\la_i(\bp)^2\phi_{\bn}+\sum_i\sum_{\bn'\prec\bp}\la_i(\bp)\ms c_{i,\bn'\bp}^0\phi_{P\bn'}
+\sum_i\sum_{\substack{\bn'\prec\bp\\ [\bn'']\preceq[\bn']}}c_{i,\bn'\bp}^0\tilde
c_{i,\bn''\bn'}^0\phi_{P\bn''}\,.
\]
Since $\bn'\prec\bp\equiv[\bn]$ if and only if $P\bn'\prec\bn$, and
$[\bn'']=[P\bn'']\preceq[\bn']\prec\bn$ implies that $P\bn''\prec\bn$, we can write the above
equality in the form
\begin{equation}
  \label{sumiJi02}
  \sum_i\big(J_i^{\,0}\big)^2\phi_{\bn}
  =\sum_i\la_i([\bn])^2\phi_{\bn}+\sum_{\bn'\prec\bn}c_{\bn'\bn}\phi_{\bn'}\,.
\end{equation}
This shows that the operator $\sum_i(J_i^{\,0})^2$ is indeed triangular in the basis $\cB$ of
$\cR_n$, with eigenvalues $\sum_i\la_i([\bn])^2$. It follows from Eqs.~\eqref{HpJs}
and~\eqref{Jmin} that $H'$ is also upper triangular in the basis $\cB$, and that its eigenvalues
are given by
\[
E_{\bn}=Nb^2-4\sum_i\la_i([\bn])^2\,,\qquad \bn\in\NN_0^N\,.
\]
The latter expression can be simplified by noting that if $[\bn]\equiv\bp=(p_1,\dots,p_N)$ and
\[
p_{k-1}>p_k=\cdots=p_{k+s}>p_{k+s+1}
\]
we have
\[
l(p_{k+j})=k\,,\qquad \#(p_{k+j})=s+1\,,\qquad 0\leq j\leq s\,,
\]
and therefore
\[
\lambda_{k+j}(\bp)=p_{k+j}+\frac12\,(1-b)+a(N+j-k-s)=
p_{k+s-j}+\frac12\,(1-b)+a(N+j-k-s)
\]
for $j=0,\dots,s$. Hence
\[
  \sum_{j=0}^s\lambda_{k+j}(\bp)^2 =
  \sum_{j=0}^s\Big(p_{k+s-j}+\frac12\,(1-b)+a(N+j-k-s)\Big)^2
  =\sum_{i=k}^{k+s}\Big(p_i+\frac12\,(1-b)+a(N-i)\Big)^2,
\]
which yields
\begin{equation}
  \label{Ebn}
  E_{\bn} = Nb^2-\sum_i\big(2p_i+1+2a(N-i)-b\big)^2\,,\qquad \bp\equiv[\bn]\,.
\end{equation}
Finally, since $\hH=\rho H'\rho^{-1}$ (cf.~Eq.~\eqref{HpJs}), the previous discussion implies that
$\hH$ is upper triangular in the basis $\rho\cB$ of $\rho\cR_n$, and that its eigenvalues are also
given by Eq.~\eqref{Ebn} with $\bn\in\NN_0^N$.

\subsection{Spectrum of $\Hsc$}

As discussed at the beginning of this section, the scalar Hamiltonian $\Hsc$
is equivalent to its extension $\tHsc$ to the symmetric space $\Lasc(L^2(\RR^N))$,
which coincides with the restriction to the latter space of the operator
$\hH$ (see Eq.~\eqref{tHs}).
Since the eigenfunctions of $\hH$ span the finite-dimensional subspace $\rho\cR_n\subset
L^2(\RR^N)$, for the purposes of computing the discrete spectrum of $\Hsc$ we can restrict
ourselves to the corresponding subspace $\Lasc(\rho\cR_n)$. We can construct a basis of the latter
space by extracting a linearly independent set from the system of generators $\Lasc(\rho\cB)$,
where $\cB$ is the basis of $\cR_n$ considered in the previous subsection. In this way we easily
obtain the basis whose elements are the functions
\[
\psi_{\bp}=\rho\,\Lasc\phi_{\bp}=\rho\,\Lasc\big(\prod_i\e^{2p_ix_i}\big),\qquad p_1\ge
p_2\ge\cdots\ge p_N,\quad p_i\in\{0,1,\dots,n\},
\]
ordered in such a way that $\psi_\bp$ precedes $\psi_{\bp'}$ whenever $\bp\prec\bp'$.
It is straightforward to show that the operator $\tHsc$ is upper triangular in the above basis,
with eigenvalues given by
\begin{equation}
  \label{Ebpsc}
  E_{\bp} = Nb^2-\sum_i\big(2p_i+1+2a(N-i)-b\big)^2\,,\qquad 0\le p_N\le\cdots\le p_1\le n\,.
\end{equation}
Indeed, taking into account that $\hH$ commutes with the symmetrizer $\Lasc$ and
acts triangularly on the basis $\rho\cB$, we have
\[
\tHsc\psi_{\bp}=\Lasc\hH(\rho\phi_{\bp})=
\Lasc\big(E_{\bp}\,\rho\phi_{\bp}+\sum_{\bn\prec\bp}c_{\bn\bp}\,\rho\phi_{\bn}\big)
=E_{\bp}\psi_{\bp}+\sum_{\bn\prec\bp}c_{\bn\bp}\,\Lasc(\rho\phi_{\bn}).
\]
Since
\[
\Lasc(\rho\phi_{\bn})=\Lasc(\rho\phi_{[\bn]})=\psi_{[\bn]}\,,
\]
we finally obtain
\[
\tHsc\psi_{\bp}=E_{\bp}\psi_{\bp}+\sum_{\bp'\prec\bp}\bigg(\sum_{\bn;\ms[\bn]=\bp'}c_{\bn\bp}\bigg)\psi_{\bp'},
\]
as claimed.

\subsection{Spectrum of $H$}

The computation of the spectrum of $H$ proceeds along the same lines. Note, first of all, that $H$
is equivalent to its extension $\tH$ to $\La(L^2(\RR^N)\otimes\Si)$, which in turn is equal to the
restriction of $\hH\otimes\id$ to the latter space. As before, in order to compute the spectrum
of $H$ we should restrict ourselves to the finite-dimensional subspace $\La(\rho\cR_n\otimes\Si)$.
A basis of the latter space is obtained by extracting a linearly independent set from the system
of generators $\La(\rho\cB\otimes\Si)$. This is easily seen to yield the spin functions
\begin{equation}\label{Psips}
\Psi_{\bp,\bs}=\rho\La(\phi_\bp\ket\bs)\,,
\end{equation}
where the quantum numbers $\bp$ and $\bs$ satisfy
\begin{subequations}\label{condss}
  \begin{align}
\text{i)}&\quad p_1\ge p_2\ge\cdots\ge p_N,\quad p_i\in\{0,1,\dots,n\},\label{conds1}\\[1mm]
\text{ii)}&\quad p_i=p_j,\quad i<j\quad\implies\quad\begin{cases}
  s_i\ge s_j,&\quad\vep=1\\
  s_i>s_j,&\quad\vep=-1\,.
\end{cases}\label{conds2}
\end{align}
\end{subequations}
The Hamiltonian $\tH$ is then upper triangular in the basis consisting of the
functions~\eqref{Psips}-\eqref{condss}, ordered so that $\Psi_{\bp,\bs}$
precedes $\Psi_{\bp',\bs'}$ whenever $\bp\prec\bp'$. Indeed, a calculation similar to
the one in the previous subsection shows that
\begin{align}
  \tH\Psi_{\bp,\bs}&=(\hH\otimes\id)\big(\rho\La(\phi_{\bp}\ket{\bs})\big)=
  \La\big(\hH(\rho\phi_{\bp})\ket\bs\big)\notag\\
  &=\La\big(E_{\bp}\,\rho\phi_{\bp}\ket\bs+\sum_{\bn\prec\bp}c_{\bn\bp}\,\rho\phi_{\bn}\ket\bs\big)
  =E_{\bp}\Psi_{\bp,\bs}+\sum_{\bn\prec\bp}c_{\bn\bp}\,\Psi_{\bn,\bs}\,.\label{tHPsi}
\end{align}  
Although a given pair of quantum numbers $(\bn,\bs)$ in the last sum of the previous equation
need not satisfy conditions~\eqref{condss}, it is easy to see that there is a permutation
$P$ (depending on both $\bn$ and $\bs$) such that $P(\bn)=[\bn]$ and
$P(\bs)=\bs'$ do satisfy~\eqref{condss}.
Since $\Psi_{\bn,\bs}$ differs from the basis vector $\Psi_{[\bn],\bs'}$ at most by a sign,
and $\bn\prec\bp$ implies that $[\bn]\prec\bp$, all the terms in the last sum of Eq.~\eqref{tHPsi}
precede $\Psi_{\bp,\bs}$. This establishes our claim and shows that the eigenvalues of
$\tH$, and thus of $H$, are the numbers
\begin{equation}
  \label{Ebpbs}
  E_{\bp,\bs} = Nb^2-\sum_i\big(2p_i+1+2a(N-i)-b\big)^2\,,
\end{equation}
where the quantum numbers $(\bp,\bs)$ satisfy conditions~\eqref{condss}.

\section{The chain's partition function}\label{sec.partfunc}

In this section we shall compute the partition function of the Frahm--Inozemtsev
chain~\eqref{cH}-\eqref{cHsites} by exploiting its relation with the dynamical models~\eqref{H}
and~\eqref{Hsc}. Indeed, we have seen in Section~\ref{sec.themodels} that if we set $b=\be a$ with
$\be>2(N-1)$, and take the limit $a\to\infty$, the eigenfunctions of $H$ become sharply peaked
around the coordinates of the minimum $\bxi=(\xi_1,\dots,\xi_N)$ of the potential $U$ in the
set~\eqref{C}. Note that the condition $\be>2(N-1)$ guarantees that the inequality~\eqref{bmin} is
fulfilled for sufficiently large $a$, so that the Hamiltonians $H$ and $\Hsc$ of the spin and
scalar dynamical models possess a non-empty point spectrum. {}From Eq.~\eqref{H-chain} and the
definition~\eqref{cH} of the Hamiltonian of the FI chain, it follows that the eigenvalues of $H$
are approximately given by
\begin{equation}\label{Eij}
E_{ij}\simeq\Esc_i+4a\,\cE_j\,,\qquad a\gg1,
\end{equation}  
where $\Esc_i$ and $\cE_j$ are two arbitrary eigenvalues of $\Hsc$ and $\cH$, respectively.
The latter formula cannot be directly used to compute the spectrum of 
the FI chain, since it is not known a priori which
eigenvalues of $H$ and $\Hsc$ combine to yield a given eigenvalue $\cE_j$ of $\cH$.
However, Eq.~\eqref{Eij} immediately yields the {\em exact} formula
\begin{equation}\label{ZZZ}
\cZ(T)=\lim_{a\to\infty}\frac{Z(4aT)}{\Zsc(4aT)}
\end{equation}
expressing the partition function $\cZ$ of the FI chain in terms of the partition functions $Z$
and $\Zsc$ of $H$ and $\Hsc$. We shall next evaluate $\cZ$ by computing the large $a$ limit of the
partition functions $Z$ and~$\Zsc$.

\subsection{Partition function of $\Hsc$}

Consider, to begin with, the partition function of the scalar model. Expanding Eq.~\eqref{Ebpsc}
in powers of $a$ we get
\begin{equation}
  \label{Ebpsca}
  E_{\bp} = E_0+4a\sum_i p_i(\be+2i-2N)+O(1)\,,
\end{equation}
where
\begin{multline*}
E_0 =N\be^2a^2-a^2\sum_i\big(\be+2i-2N\big)^2
+2a\sum_i\big(\be+2i-2N\big)\\
=\frac23\,aN\big(a(N-1)(3\be-2N+1)+3(\be-N+1)\big)
\end{multline*}
is a constant independent of $\bp$. {}From now on we shall subtract from both $\Hsc$ and $H$
the constant energy $E_0$. With this proviso, when $a$ is sufficiently large
the partition function $\Zsc$ is approximately given by
\[
\Zsc(4aT)\simeq\sum_{0\le p_N\le\cdots\le p_1\le n}\prod_iq^{p_i(\be+2i-2N)}\,.
\]
Note that, by condition~\eqref{cond}, the coefficient of $p_i$ in the RHS
of the previous formula is strictly positive. In terms of the new summation
indices
\[
n_i=p_i-p_{i+1}\,,\qquad 1\leq i\leq N\quad (p_{N+1}\equiv0)\,,
\]
we have $p_i=\smash{\sum\limits_{j=i}^N}n_j$, so that
\[
\sum_ip_i(\be+2i-2N)=\sum_{j\geq i}
n_j(\be+2i-2N)=\sum_{j=1}^Nn_j\sum_{i=1}^j(\be+2i-2N)
=\sum_{j=1}^Njn_j(\be-2N+j+1)\,.
\]
Hence
\[
\Zsc(4aT)\simeq\sum_{\substack{n_1,\dots,n_N\ge0\\n_1+\cdots+n_N\leq n
}}\prod_jq^{jn_j(\be-2N+j+1)}\,,
\]
where again the coefficient of $n_j$ is strictly positive on account of~\eqref{cond}. For finite
$a$, it is not easy to evaluate in closed form the sum in the RHS of the previous equation due to
the restriction $n_1+\cdots+n_N\le n$. This restriction is in fact another peculiarity of the
present (hyperbolic) model, not present in the trigonometric case~\cite{FG05}. However, since
$n\to\infty$ as $a\to\infty$ on account of Eq.~\eqref{nmax}, taking into account that $q<1$ we
finally have
\begin{align}
\lim_{a\to\infty}\Zsc(4aT)&=\sum_{n_1,\dots,n_N\geq0}\prod_jq^{jn_j(\be-2N+j+1)}
=\prod_j\sum_{n_j=0}^\infty q^{jn_j(\be-2N+j+1)}\notag\\
\label{Zsc}
&=\prod_j\big(1-q^{\cF(j)}\big)^{-1}\,,
\end{align}
where the \emph{dispersion relation} $\cF(j)$ is defined by
\begin{equation}\label{cF}
\cF(j)=j(\be-2N+j+1)\,.
\end{equation}

\subsection{Partition function of $H$}

By Eq.~\eqref{Ebpbs}, the partition function of the spin dynamical model~\eqref{H} can be written as
\[
Z(4aT)=\sum_{0\le p_N\le\cdots\le p_1\le n}d_{\bp}\,q^{\frac{E_\bp}{4a}}\,,
\]
where $E_\bp$ is given by Eq.~\eqref{Ebpsc} and $d_\bp$ is the number of spin quantum numbers
$\bs$ satisfying condition~\eqref{conds2}. Writing the quantum number $\bp$ as
\begin{equation}
\bp=\big(\overbrace{\vphantom{1}\nu_1,\dots,\nu_1}^{k_1},\dots,
\overbrace{\vphantom{1}\nu_r,\dots,\nu_r}^{k_r}\big),\qquad 0\le\nu_r<\cdots<\nu_1\le n\,,
\label{bp}
\end{equation}
we easily obtain
\begin{equation}\label{dp}
  d_\bp=d(\bk)\equiv\begin{cases}
    \prod\limits_{i=1}^r\binom{m+k_i-1}{k_i}\,,&\quad\vep=1\,,\\[5mm]
\prod\limits_{i=1}^r\binom{m}{k_i}\,,&\quad\vep=-1\,.
\end{cases}
\end{equation}
Using the asymptotic expansion~\eqref{Ebpsca} and ignoring (as in the scalar case)
the constant energy $E_0$, we have
\[
Z(4aT)\simeq\sum_{0\le p_N\le\cdots\le p_1\le n}d_{\bp}\,q^{\sum\limits_i p_i(\be+2i-2N)}\,.
\]
Setting
\begin{equation}\label{Ki}
K_i = \sum_{j=1}^ik_j
\end{equation}
and using Eq.~\eqref{bp} we have
\[
  \sum_ip_i(\be+2i-2N)=\sum_{j=1}^r\nu_j\sum_{i=K_{j-1}+1}^{K_j}(\be+2i-2N)
  =\sum_{j=1}^r\nu_jk_j\big(\be-2N+2K_j-k_j+1\big)\,.
\]
Introducing, as before, the variables
\[
n_i=\nu_i-\nu_{i+1}\,,\qquad 1\leq i\leq r\quad (n_r\equiv \nu_r)\,,
\]
in terms of which $\nu_j=\sum\limits_{i=j}^rn_i$, we can write
\[
\sum_ip_i(\be+2i-2N)=\sum_{1\leq j\leq i\leq r}n_ik_j\big(\be-2N+2K_j-k_j+1\big)
\equiv  \sum_{i=1}^rn_iN_i\,,
\]
with
\begin{align}
  N_i &= \sum_{j=1}^ik_j(\be-2N+2K_j-k_j+1)
  =(\be-2N+1)K_i-\sum_{j=1}^ik_j^2+2\sum_{1\leq
    l\leq j\leq i}k_jk_l\notag\\
  &= (\be-2N+1)K_i+\sum_{j=1}^ik_j^2+\sum_{1\leq l\neq j\leq i}k_jk_l=\cF(K_i)\,,
  \label{Ni}
\end{align}
where $\cF$ is defined in Eq.~\eqref{cF}. We thus have
\[
Z(4aT)\simeq\sum_{\bk\in\cP_N}d(\bk)\sum_{\substack{n_1,\dots,n_{r-1}>0,n_r\geq0\\
  n_1+\cdots+n_r\leq n}}\prod_iq^{n_i\cF(K_i)}\,,
\]
where $\cP_N$ is the set of all partitions of $N$ with order taken into account. Since
$\cF(K_i)>0$ for all $i$, and $n\to\infty$ as $a\to\infty$, we finally obtain
\begin{align}
  \lim_{a\to\infty}Z(4aT)&=\sum_{\bk\in\cP_N}d(\bk)\sum_{n_1,\dots,n_{r-1}>0}
  \sum_{n_r\geq0}\prod_iq^{n_i\cF(K_i)}
  = \sum_{\bk\in\cP_N}d(\bk)\sum_{n_r=0}^\infty
  q^{n_r\cF(K_r)}\prod_{i=1}^{r-1}\sum_{n_i=1}^\infty q^{n_i\cF(K_i)} \notag \\\label{Z} &=
  \sum_{\bk\in\cP_N}d(\bk)\big(1-q^{\cF(K_r)}\big)^{-1}\prod_{i=1}^{r-1}\frac{q^{\cF(K_i)}}{1-q^{\cF(K_i)}}\,.
\end{align}

\subsection{Partition function of the FI chain}

Substituting Eqs.~\eqref{Zsc} and~\eqref{Z} into the freezing trick relation~\eqref{ZZZ} we obtain
the following closed formula for the partition function of the FI chain:
\begin{equation}
  \label{cZ}
  \cZ(T) = \sum_{\bk\in\cP_N}d(\bk)\,
  q^{\sum_{i=1}^{r-1}\cF(K_i)}
  \prod_{i=1}^{N-r} (1-q^{\cF(K'_i)})\,,
\end{equation}
where $K_i$ is given by Eq.~\eqref{Ki}, $r$ is the number of components of $\bk$, and
\[
\{K'_1,\dots,K'_{N-r}\}=\{1,\dots,N\}\setminus\{K_1,\dots,K_r\}\,.
\]
In fact, since $K_r=\sum\limits_{i=1}^{r}k_i=N$ we have
\[
\{K'_1,\dots,K'_{N-r}\}=\{1,\dots,N-1\}\setminus\{K_1,\dots,K_{r-1}\}\,.
\]
It should be noted that the expression~\eqref{cZ} for the partition function of the FI chain
coincides with the corresponding ones for the HS~\cite{FG05} and PF~\cite{BBS08} chains, provided
that the dispersion relation~\eqref{cF} is replaced by
\[
\cF_{\mathrm{HS}}(j)=j(N-j)\,,\qquad \cF_{\mathrm{PF}}(j)=j\,.
\]
{}From Eq.~\eqref{cZ} it readily follows that the energies of the FI chain are of the form
\begin{equation}\label{cEde}
\cE(\bde)=\sum_{i=1}^{N-1}\de_i\cF(i)\,,
\end{equation}
where $\bde=(\de_1\cdots\de_{N-1})$ and $\de_i\in\{0,1\}$. In fact, Eq.~\eqref{cZ} and the
discussion in~\cite{BBS08} imply that the numbers $\de_i$ are the well-known {\em motifs}
introduced in Ref.~\cite{HHTBP92}. More precisely, from Refs.~\cite{BBHS07,BBS08} it follows that
the motifs (with their respective multiplicities taken into account) of the ferromagnetic
($\vep=1$) su($m)$ chain are generated by setting
\begin{equation}
  \label{motifsum}
  \de_i=\begin{cases}1\,,& \ka_{i+1}>\ka_i\\ 0\,, & \ka_{i+1}\le \ka_i\,,
    \end{cases}
\end{equation}
where the numbers $\ka_i$ ($i=1,\dots,N$) are independent and take the values $0,1,\dots,{m-1}$. In
particular, in the case of spin $1/2$ ($m=2$) Eq.~\eqref{motifsum} is equivalent to
\begin{equation}\label{motifsum2}
\de_i=\ka_{i+1}(1-\ka_i)\,,\qquad \ka_i\in\{0,1\}\,,
\end{equation}
which together with Eq.~\eqref{cEde} yields the empiric formula for the spectrum proposed in
Ref.~\cite{FI94}.

We shall finish this section by finding the relation between the spectra of the ferromagnetic and
antiferromagnetic FI chains. {}From Eqs.~\eqref{cH} and~\eqref{cHz} it follows that
\begin{equation}\label{cHs}
\cH^{\mathrm{F}}+\cH^{\mathrm{AF}}=2\sum_{i\ne j}h_{ij}\,,
\end{equation}
with
\begin{equation}\label{hij}
h_{ij}=\frac14\sinh^{-2}(\xi_i-\xi_j)=\frac{\ze_i\ze_j}{(\ze_i-\ze_j)^2}\,.
\end{equation}
The RHS of Eq.~\eqref{cHs} is the maximum energy of the ferromagnetic chain when $m\ge N$,
corresponding to states completely antisymmetric under spin permutations. This maximum
energy can be easily computed from Eq.~\eqref{cEde} noting that
$\cF(i)>0$ for all $i$, and that the motif $\bde=(1\,1\cdots1)$ is
compatible with~\eqref{motifsum} when $m\ge N$ (take, for instance, $\ka_i=i$).
We thus have
\begin{equation}\label{cEmaxF}
2\sum_{i\ne j}h_{ij}=\sum_{i=1}^{N-1}\cF(i)=\frac16\,N(N-1)(3\be-4N+2)\,.
\end{equation}
From Eqs.~\eqref{cEde}, \eqref{cHs} and~\eqref{cEmaxF} it also follows that if
$(\de_1\cdots\de_{N-1})$ is a motif for the ferromagnetic chain then
$(1-\de_1\,\cdots\,1-\de_{N-1})$ is a motif for the antiferromagnetic one, and vice versa. This
property, empirically discovered by Haldane~\cite{Ha93} for the original HS chain, is a
manifestation of the boson-fermion duality recently established in Ref.~\cite{BBHS07}.

\section{The chain's spectrum}\label{sec.spectrum}

Over the last few years, there has been growing evidence of the singular character of spin chains
of Haldane--Shastry type in connection with a number of well-known tests of integrability versus
chaos in quantum systems. In this section we shall analyze whether the FI chain~\eqref{cH}
---which is most likely integrable~\cite{FI94}--- behaves in this respect as other integrable
chains of HS type previously studied in the literature. To this end, we shall take advantage of
the explicit formula~\eqref{cZ} for the partition function to compute the spectrum for relatively
large values of $N$ and fixed $m=2,3,\dots$.

To begin with, on account of Eqs.~\eqref{cHs} and \eqref{cEmaxF} we can restrict ourselves to
studying either the ferromagnetic or the antiferromagnetic chain~\eqref{cH}. Due to the form of
the degeneracy factors~\eqref{dp} appearing in the partition function, it is far more convenient to
deal with the latter chain (corresponding to $\vep=-1$), as we shall do in the rest of this
section. In the first place, it is clear from Eqs.~\eqref{cF} and~\eqref{cEde} that when
$\be$ is sufficiently large the energy levels are split into disjoint ``clusters'' centered around
the numbers
\begin{equation}\label{centers}
\be\sum_i\de_i\,i\,,
\end{equation}  
where the components of the (antiferromagnetic) motif $\bde$ are defined in terms of the
independent numbers $\ka_i\in\{0,1,\dots,m-1\}$ as
\begin{equation}
  \label{motifsumanti}
  \de_i=\begin{cases}0\,,& \ka_{i+1}>\ka_i\\ 1\,, & \ka_{i+1}\le \ka_i\,.
    \end{cases}
\end{equation}
It is straightforward to show that the centers of the clusters~\eqref{centers} are the numbers
$\be j$, where $j$ is an integer ranging from $\frac12N'(2N-m-mN')$ to $\frac12N(N-1)$,
$N'\equiv\lfloor N/m\rfloor$ denoting the integer part of $N/m$. Indeed, the
minimum value of $j$ corresponds to the motif
\begin{equation}
\big(\overbrace{\vphantom{|}0\cdots0}^{m'}\:
\overbrace{\vphantom{|}1\,0\cdots0}^{m}\,\cdots\,
\overbrace{\vphantom{|}1\,0\cdots0}^{m}\big),
\qquad m'\equiv (N-1)\mod m\,,\label{motifmin}
\end{equation}
while the maximum value is obtained from the motif $(1\,1\cdots 1)$.

\begin{figure}[h]
\includegraphics[height=4.3cm]{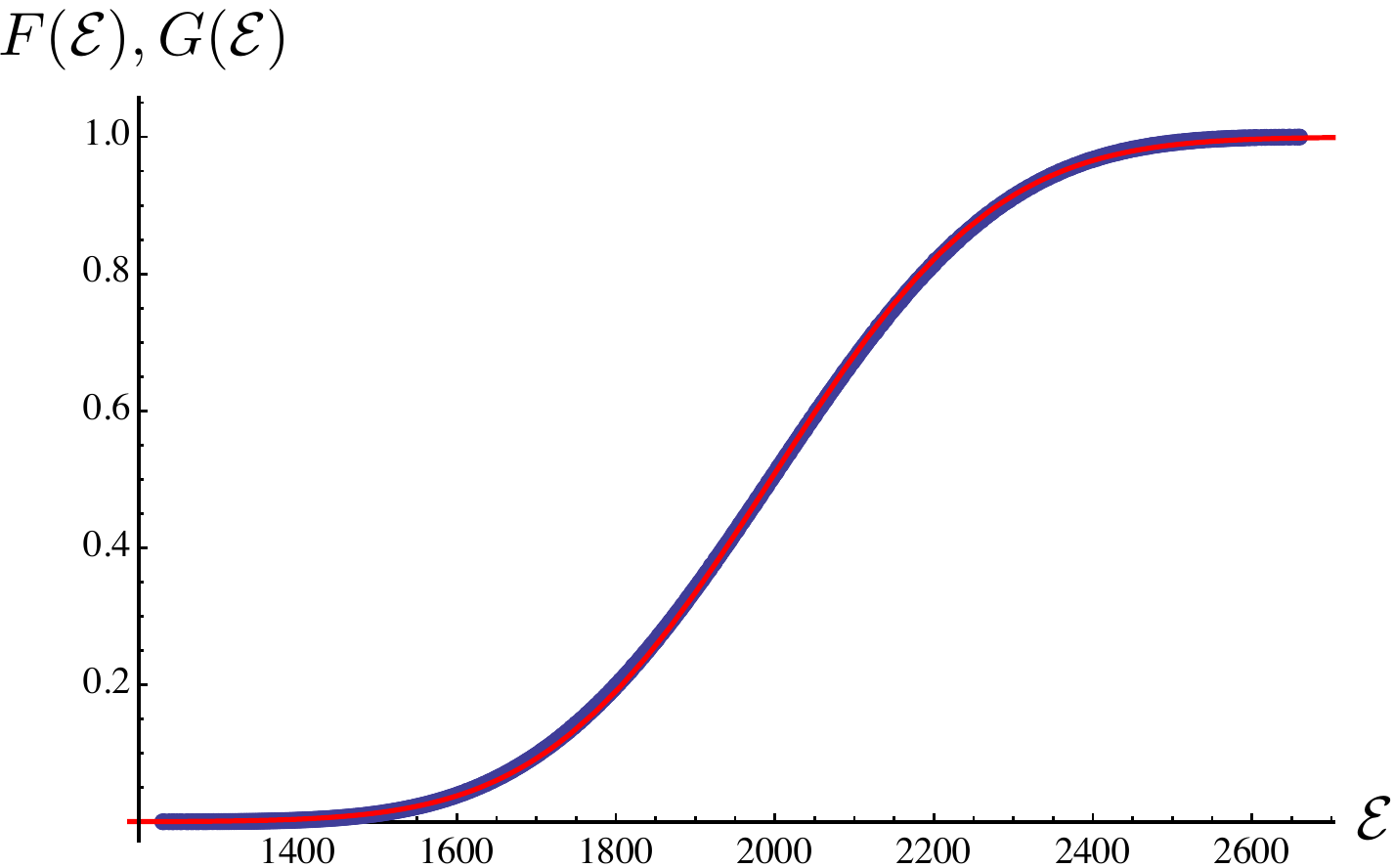}\hfill
\includegraphics[height=4.3cm]{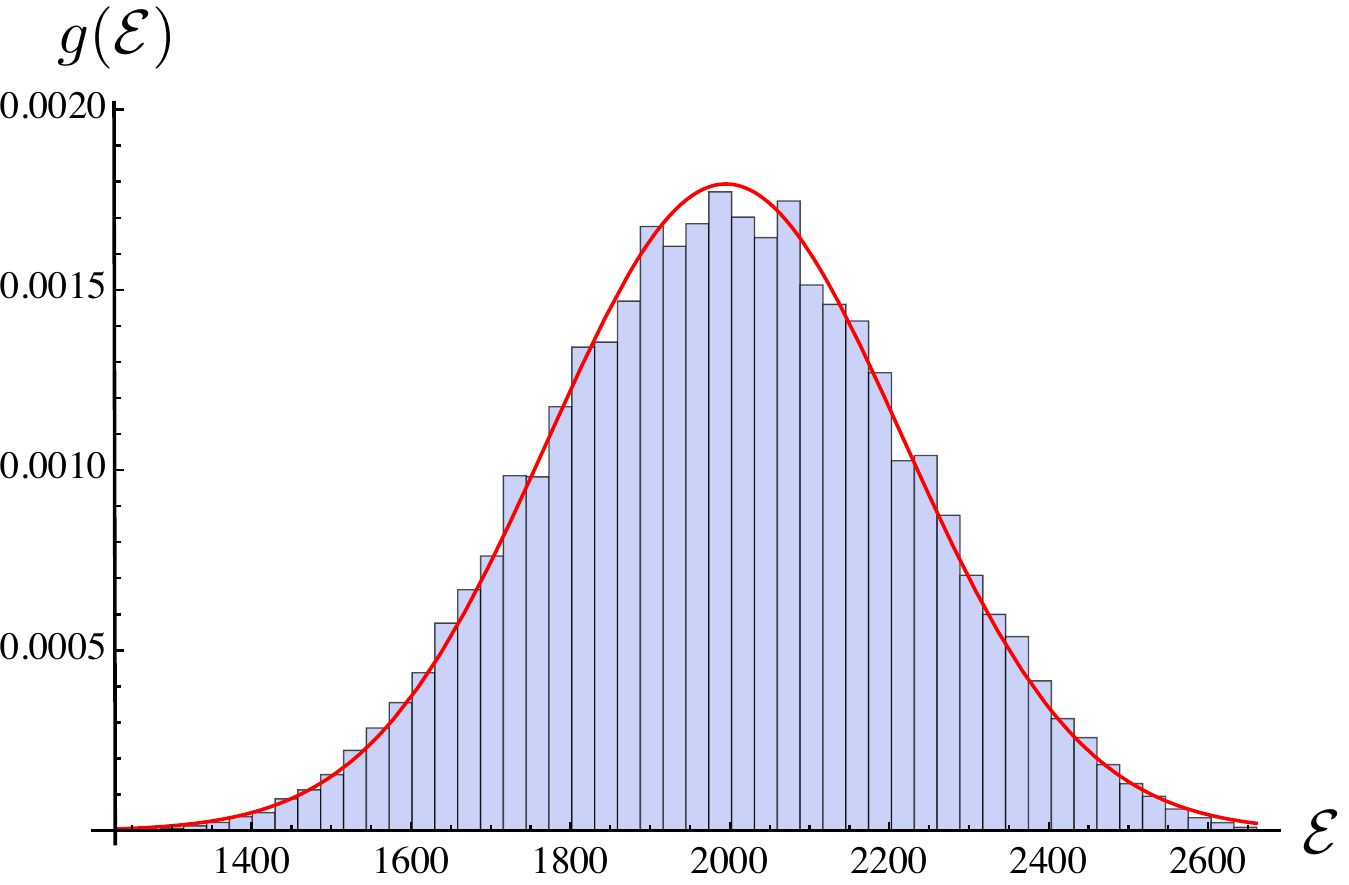}
\caption{Left: plot of the cumulative Gaussian distribution~\eqref{GcE} (continuous red line)
  vs.~the cumulative level density~\eqref{FcE} (blue dots) for the antiferromagnetic FI chain with
  $N=20$, $m=2$, $\be=40$. Right: histogram of the density of energy levels (normalized to unity)
  of the latter chain vs.~the Gaussian distribution
  $g(\cE)=\frac1{\sqrt{2\pi}\,\si}\e^{-\frac{(\cE-\mu)^2}{2\si^2}}$ (continuous red line).}
\label{fig.levden}
\end{figure}

In order to avoid the splitting of the spectrum just described, we shall henceforth assume that
$\be$ is $O(N)$ (cf.~Eq.~\eqref{cond}). With this proviso, we have computed the spectrum of the
antiferromagnetic FI chain for several values of $N$, $m$ and $\be$ using the formula~\eqref{cZ}
for the partition function. The first statistical property that we have analyzed (which
is essential for the process of ``unfolding'' the spectrum, as described below) is the
cumulative level density (normalized to unity)
\begin{equation}\label{FcE}
F(\cE)=m^{-N}\sum_{i;\cE_i\le\cE}d_i\,,
\end{equation}
where $d_i$ is the degeneracy of the energy $\cE_i$. It is apparent from our results that, for
even moderately large values of $N$, $F(\cE)$ is very well
approximated by the cumulative Gaussian law
\begin{equation}\label{GcE}
  G(\cE)\equiv\frac1{\sqrt{2\pi}\,\si}\int_{-\infty}^\cE\e^{-\frac{(\cE'-\mu)^2}{2\si^2}}\diff\cE'
  = \frac12\bigg[1+\erf\bigg(\frac{\cE-\mu}{\sqrt2\,\si}\bigg)\bigg]
\end{equation}
with parameters $\mu$ and $\si$ respectively equal to the mean and standard deviation of the
energy (see Fig.~\ref{fig.levden}, left). These parameters can be computed in closed form
essentially by taking traces of appropriate powers of the Hamiltonian~\eqref{cH}, with the result
(see Appendix~\ref{sec.musigma} for the details):
\begin{subequations}\label{musi}
\begin{align}
  \mu&=\frac1{12}\,\Big(1+\frac1m\Big)N(N-1)(3\be-4N+2)\label{mufinal}\,,\\[1mm]
  \si^2&=\frac{1}{360}\Big(1-\frac1{m^2}\Big)N(N-1)\big[16N^3-N^2(25\be-6)\notag\\
  &\hphantom{=\frac{1}{360}\Big(1-\frac1{m^2}\Big)N(}+N(10\be^2-35\be+26)+(5\be-6)(5\be+4)\big]\,.\label{si2final}
\end{align}
\end{subequations}
We have also observed that for $\be=O(N)$
the level density itself tends to a Gaussian distribution when the number of particles tends to
infinity, as is the case for other spin chains of HS type whose spectrum is
equispaced~\cite{BFGR09,BFG09}; see Fig.~\ref{fig.levden}, right.

We shall consider next the distribution of the spacings between consecutive levels in the
``unfolded'' spectrum. Recall~\cite{GMW98} that in order to compare different spectra in a
consistent way, it is necessary to apply to the energy levels
$\cEmin\equiv\cE_1<\dots<\cE_{L+1}\equiv\cEmax$ the unfolding mapping
$\cE_i\mapsto\eta_i\equiv\eta(\cE_i)$, where $\eta(\cE)$ is a continuous approximation to the
cumulative level density $F(\cE)$. Indeed, it can be shown that the resulting ``unfolded''
spectrum $\{\eta_i\}_{i=1}^{L+1}$ is uniformly distributed regardless of the initial level
density. In our case, by the above discussion we can take $\eta(\cE)$ as the cumulative Gaussian
density~\eqref{GcE}, i.e.,
\begin{equation}\label{eta}
\eta_i=\frac12\,
\Big[
1+\erf\Big(\frac{\cE_i-\mu}{\sqrt 2\si}\Big)
\Big]\,.
\end{equation}
By convenience, we shall consider the normalized spacings
\[
s_i=\frac{\eta_{i+1}-\eta_{i}}{L(\eta_{L+1}-\eta_1)}\,,\qquad i=1,\dots,L\,,
\]
so that $\{s_i\}_{i=1}^{L}$ has unit mean. The distribution of the spacings $s_i$ is a widely used
indicator of the integrable vs.~chaotic character of a quantum system. More precisely, the
celebrated conjecture of Berry and Tabor posits that for a ``typical'' quantum integrable system
the probability density $p(s)$ of the normalized spacings $s_i$ should be given by Poisson's law
$p(s)=\e^{-s}$. By contrast, according to the Bohigas--Giannoni--Schmidt conjecture~\cite{BGS84},
the spacings distribution of a quantum system whose classical counterpart is chaotic should
instead follow Wigner's law $p(s)=(\pi s/2)\ms\exp(-\pi s^2/4)$, as is the case for the Gaussian
orthogonal ensemble in random matrix theory~\cite{GMW98}. Interestingly, the spacings distribution
$p(s)$ of all the (integrable) spin chains of HS type studied so
far~\cite{FG05,BB06,BFGR08,BFGR08epl,BFGR09,BFG09} follows neither Poisson's nor Wigner's law.
More precisely, for these chains the cumulative spacings distribution
\[
P(s)\equiv\int_0^s p(s')\diff s'
\]
can be estimated with remarkable accuracy by the formula 
\begin{equation}\label{P}
P(s)\simeq 1-\frac{2}{\sqrt\pi\,\smax}\,\sqrt{\log\Big(\frac{\smax}s\Big)}\,,
\end{equation}
where the parameter
\begin{equation}\label{smax}
\smax\equiv\frac{\cE_{\mathrm{max}}-\cE_{\mathrm{min}}}{\sqrt{2\pi}\,\si}
\end{equation}
is approximately equal to the maximum spacing. In fact, in Refs.~\cite{BFGR08,BFGR09} we have
shown that the approximation~\eqref{P}-\eqref{smax} is valid\footnote{More precisely~\cite{BFGR09},
  Eq.~\eqref{P} holds in the range $s_0\le s\le\smax$, where
  $s_0\equiv\smax\e^{-\frac\pi4\ms\smax^2}\ll\smax$.} for any spectrum satisfying
the following conditions:

{\leftskip1.3cm\parindent=0pt\setcounter{ex}{0}\parskip=6pt%
\cond The energies are \emph{equispaced}, i.e., $\cE_{i+1}-\cE_{i}=\de\cE$ for $i=1,\dots,L$.

\cond The cumulative level density (normalized to unity) is approximately given by the Gaussian law~\eqref{GcE}.

\cond $\cE_{\mathrm{max}}-\mu\,,\,\mu-\cE_{\mathrm{min}}\gg\si$, where $\mu$ and $\si$ are the mean
and standard deviation of the spectrum.

\cond $\cE_{\mathrm{min}}$ and $\cE_{\mathrm{max}}$ are approximately symmetric with respect to
$\mu$, namely
$\vert\cE_{\mathrm{min}}+\cE_{\mathrm{max}}-2\mu\vert\ll\cE_{\mathrm{max}}-\cE_{\mathrm{min}}$.

}

In our case, we have already mentioned that when $\be=O(N)$ the FI chain satisfies condition (ii)
above. It is also easy to check that the third and fourth condition are also satisfied.
Indeed, from Eqs.~\eqref{cH}, \eqref{hij} and~\eqref{cEmaxF} it immediately follows that
\begin{equation}\label{cEmax}
\cEmax=\frac16\,N(N-1)(3\be-4N+2)\,.
\end{equation}
On the other hand, the minimum energy can be computed from Eqs.~\eqref{cF} and \eqref{cEde}
using the motif~\eqref{motifmin}, with the result
\begin{equation}\label{cEmin}
  \cEmin=\frac{N'}6\,\big[m^2 (N'+1)(2N'+1)-3 m (N'+1) (\be+1)+6 N (\be-N+1)\big]\,.
\end{equation}
Since $N'=N/m+O(1)$ and we are assuming that $\be=O(N)$, the previous equation yields
\begin{equation}\label{cEmin2}
 \cEmin=\frac{N^2}{6m}\,(3\be-4N)+O(N^2)\,.
\end{equation}
{}From Eqs.~\eqref{mufinal}, \eqref{cEmax} and~\eqref{cEmin2} 
it immediately follows that
\begin{equation}\label{maxmumin}
  \cEmax-\mu\,,\,\mu-\cEmin=\frac1{12}\Big(1-\frac1m\Big)N^2(3\be-4N)+O(N^2)\,,
\end{equation}
and thus $\cEmax-\cEmin=O(N^3)$, so that condition (iv) is clearly satisfied.
As to condition (iii), it suffices to note that $\si=O(N^{5/2})$ by Eq.~\eqref{si2final}.

\begin{figure}[H]
\includegraphics[height=4.3cm]{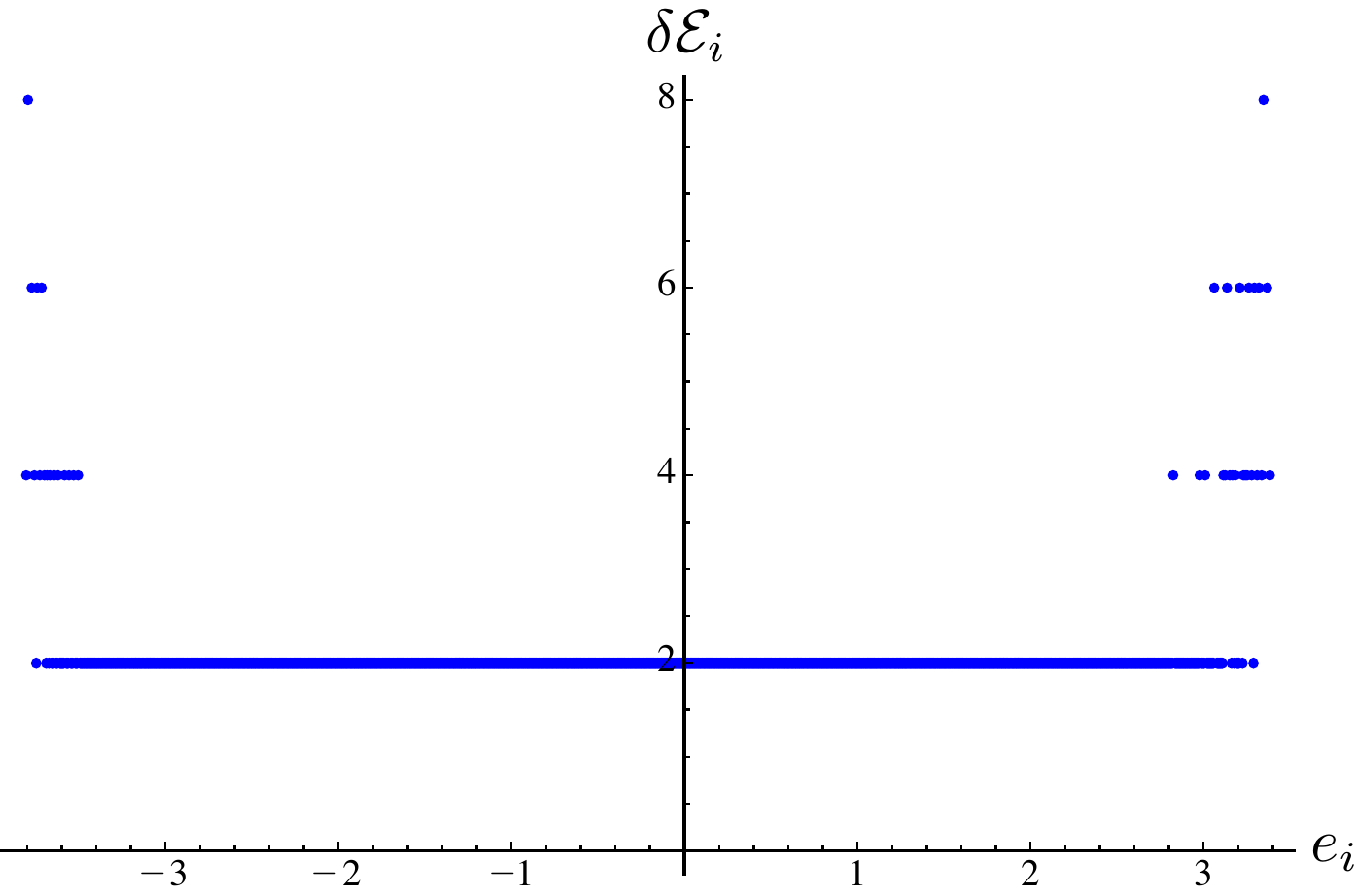}\hfill
\includegraphics[height=4.3cm]{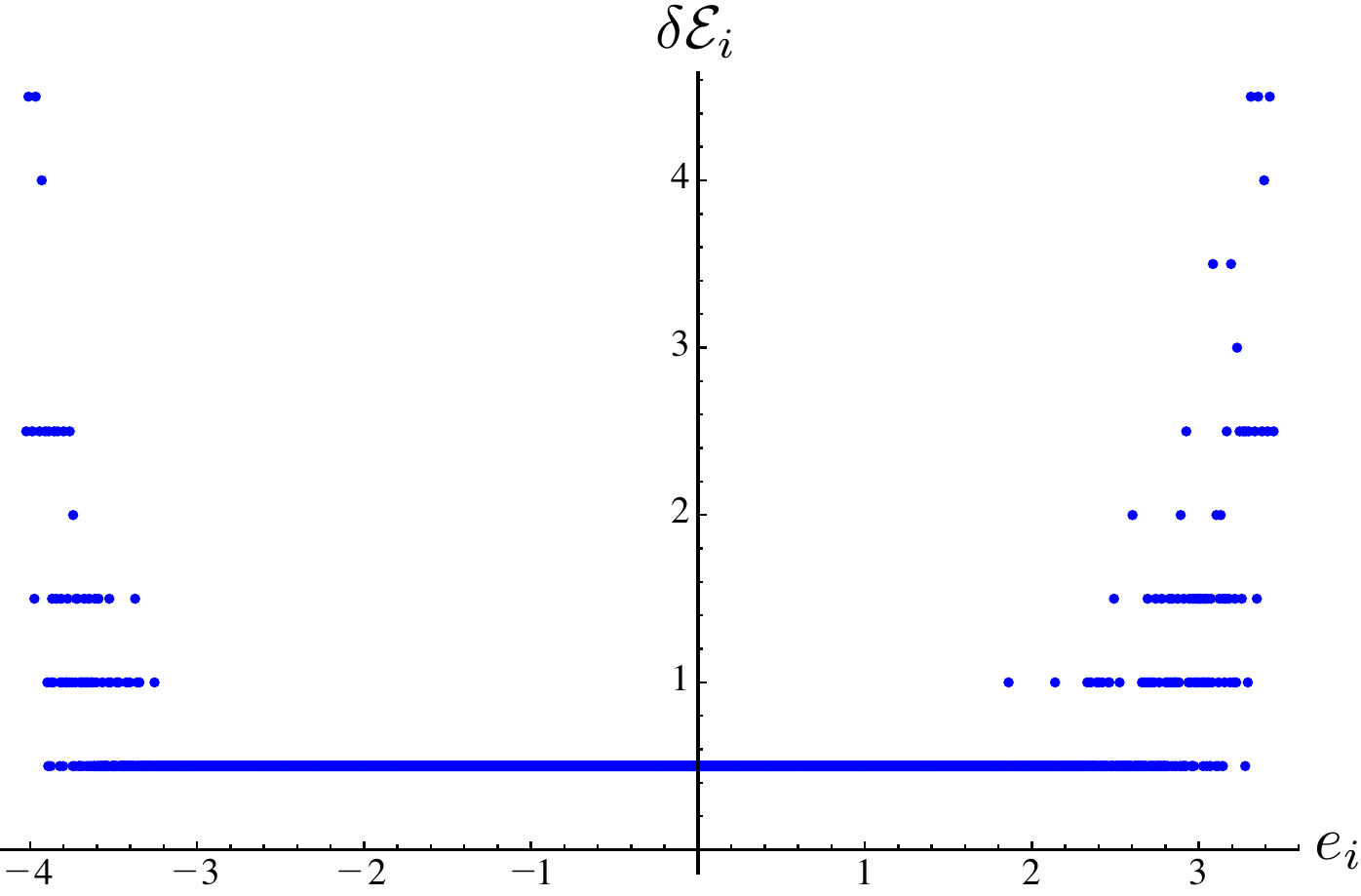}
\caption{Differences $\de\cE_i$ versus $e_i\equiv(\cE_i-\mu)/\si$ for the
antiferromagnetic FI chain with $N=24$, $m=2$ and $\be=50$ (left) and $N=18$, $m=3$ and $\be=36.5$
(right).
  \label{fig.deltaEs}}
\end{figure}

\begin{figure}[H]
  \centering
  \includegraphics[width=8cm]{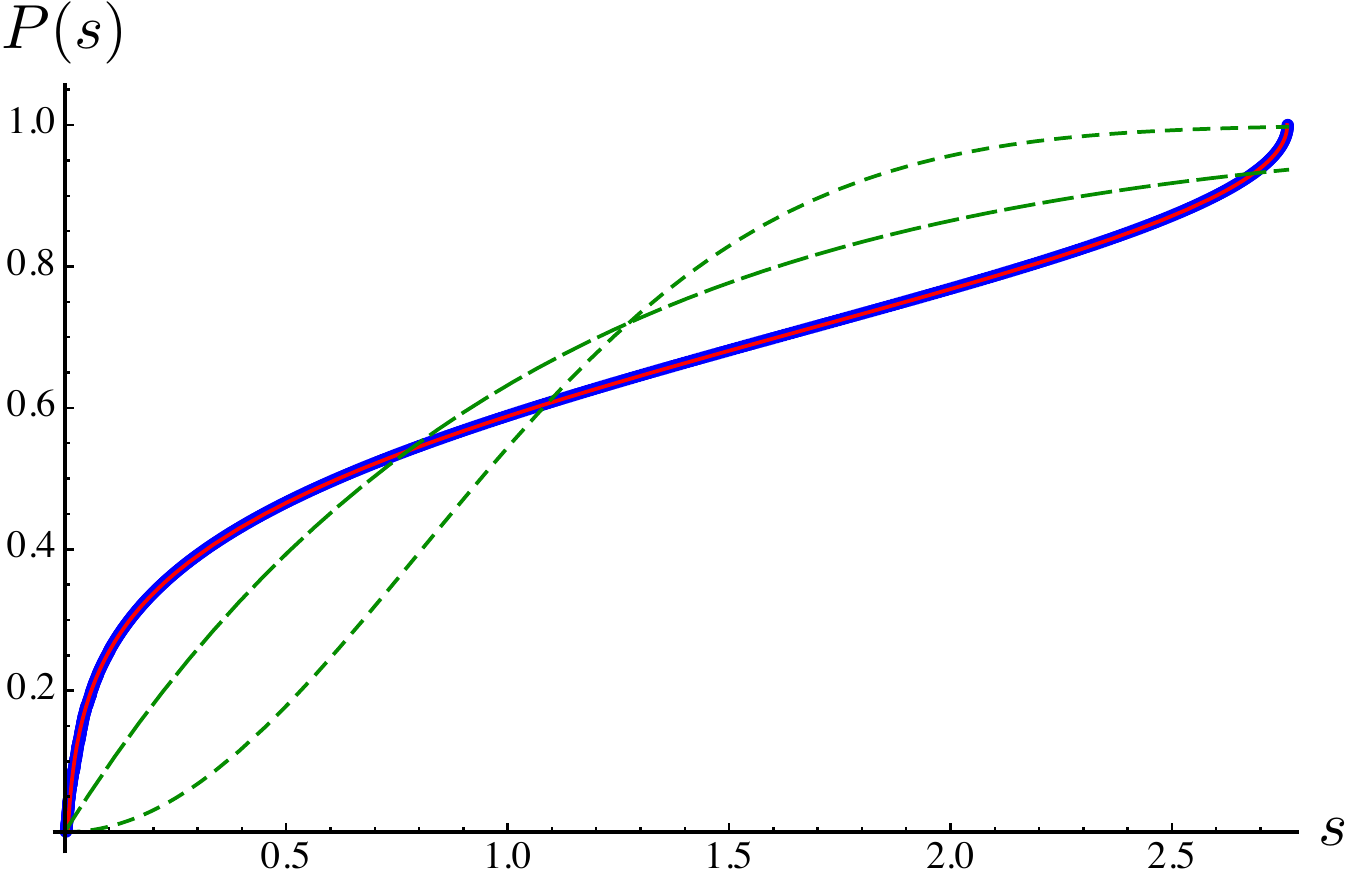}
\caption{Cumulative spacings distribution and its analytic approximation~\eqref{P}
(continuous red line) for the antiferromagnetic FI chain with $N=24$, $m=2$ and $\be=50$. For
convenience, we have also represented Poisson's (green, long dashes) and Wigner's (green, short
dashes) cumulative distributions.
  \label{fig.spacings}}
\end{figure}

Let us now examine the first condition listed above. The analysis is complicated by the fact that
the spectrum of the FI chain depends on an essential parameter $\be>2(N-1)$, in contrast to all
the chains of HS type whose spacing distribution has been studied so far. According to our
numerical computations, there are clearly two different regimes. Indeed, if $\be$ is an integer or
a rational number with a ``small'' denominator, the vast majority of the differences
$\de\cE_i\equiv\cE_{i+1}-\cE_{i}$ are equal to a single value $\de\cE$ which depends on $\be$
(e.g., $\de\cE=1$ when $\be$ is an odd integer, and $\de\cE=2$ when $\be$ is an even integer).
Thus, in this case the spectrum of the FI chain is approximately equispaced when $N$ is
sufficiently large. Moreover, our computations indicate that in this regime the differences
$\de\cE_i\ne\de\cE$ are concentrated in the tails of the level density
(see~Fig.~\ref{fig.deltaEs}). As shown in Ref.~\cite{BFGR08epl}, these two properties
together with conditions (ii)--(iv) above are sufficient to guarantee the validity
of the approximation~\eqref{P}. We have verified that this is indeed the case for a wide range of
values of $N$, $m$ and $\be$; see e.g.~Fig.~\ref{fig.spacings} for the case $N=24$, $m=2$
and $\be=50$.


\begin{figure}[h]
\includegraphics[height=4.3cm]{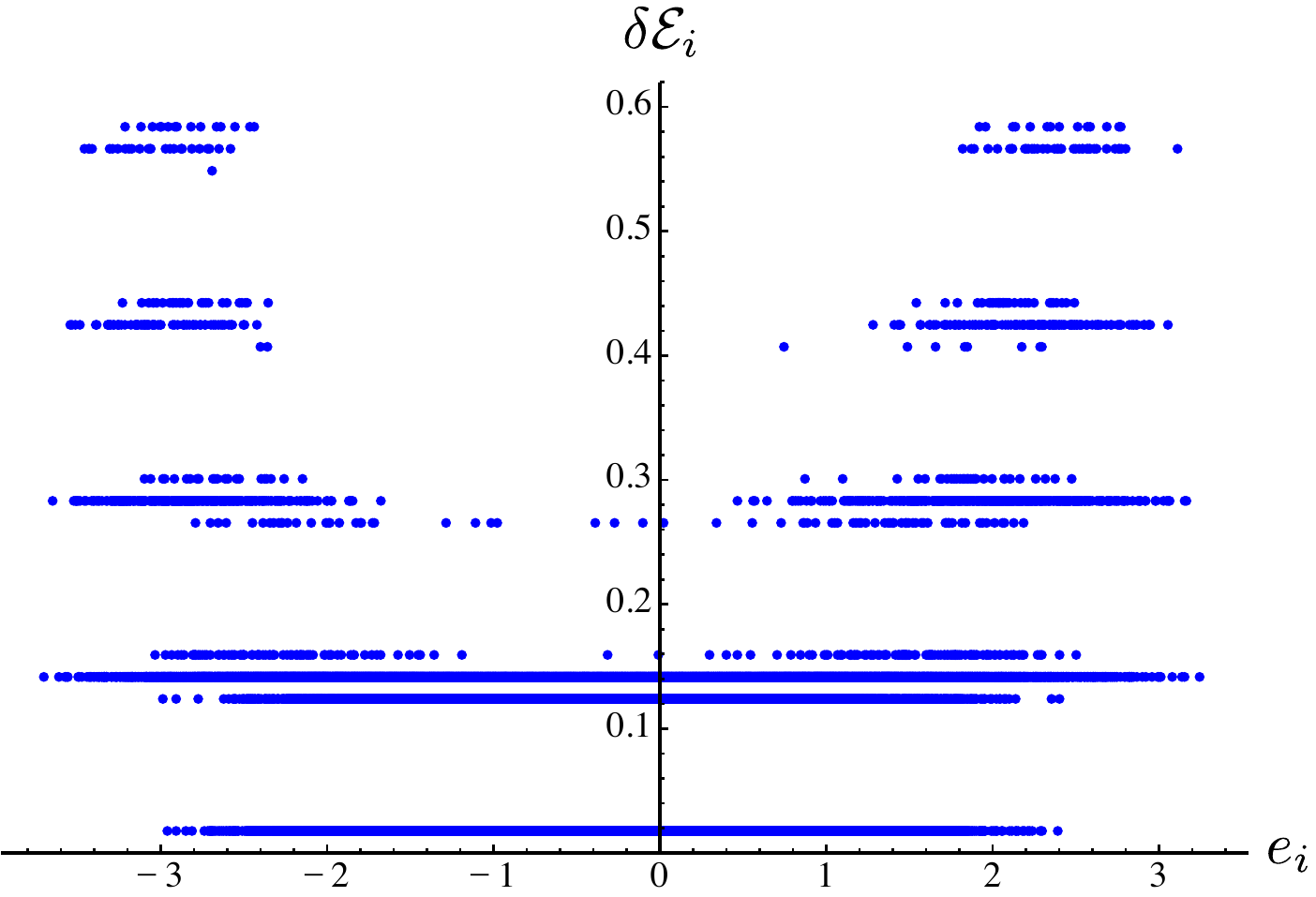}\hfill
\includegraphics[height=4.3cm]{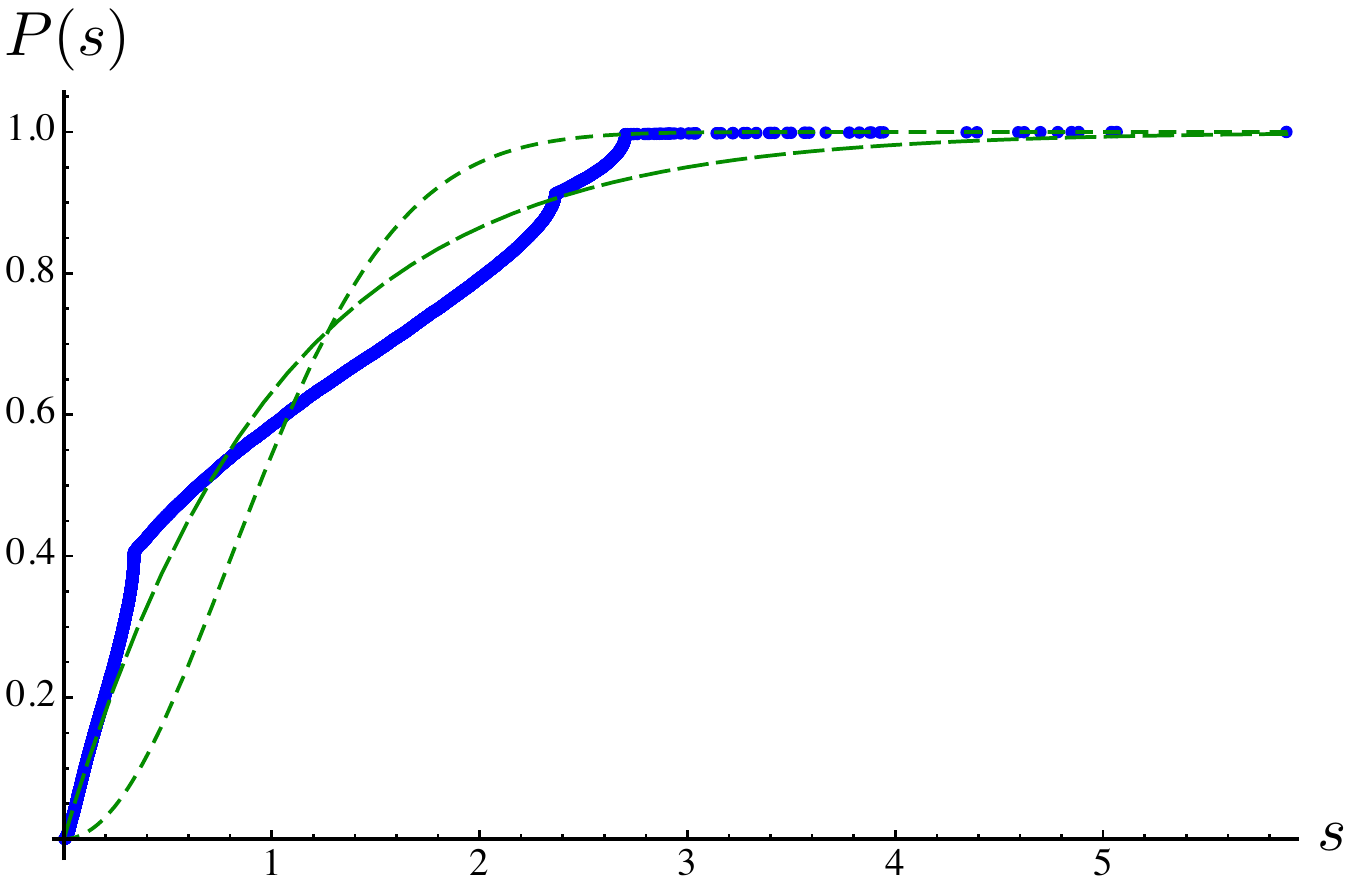}
\caption{Left: differences $\de\cE_i$ versus $e_i\equiv(\cE_i-\mu)/\si$ for the antiferromagnetic
  FI chain with $N=24$, $m=2$ and $\be=47+\pi$ (for convenience we have only shown differences
  $\de\cE_i<0.6$, which account for over 98\% of the total). Right: Cumulative spacings
  distribution for the latter chain, compared with Poisson's (green, long dashes) and Wigner's
  (green, short dashes) cumulative distributions.
  \label{fig.irrational}}
\end{figure}

Let us next examine the alternative regime in which the parameter $\be=O(N)$ is either irrational
or a rational with a non-small denominator (greater than $3$, say). The key difference with the
previous case is that the (raw) spectrum is not even approximately equispaced, so that condition
(i) does not hold. This is essentially due to the fact that the energies $\cE_i$ are of the form
$\be j+k$, with $j,k$ integers (cf.~Eqs.~\eqref{cF} and~\eqref{cEde}). As a consequence,
among the differences $\de\cE_i$ there is no single dominant value, but rather a set of several
most frequent values (cf.~Fig.~\ref{fig.irrational}, left).

Our numerical calculations suggest that the presence of several dominant energy differences gives
rise to discontinuities in the spacings distribution $p(s)$, evidenced by the appearance of
``cusps'' in the plot of $P(s)$ (see Fig.~\ref{fig.irrational}, right). The derivation of an
analytic approximation to $P(s)$ akin to Eq.~\eqref{P} will appear in a forthcoming paper.

\section{Concluding remarks}\label{sec.rems}

In this paper we have studied the hyperbolic $\mathrm{su}(m)$ spin CS model of $A_{N-1}$ type
introduced by Inozemtsev in Ref.~\cite{In96} and its associated spin chain of HS type known as the
Frahm--Inozemtsev chain~\cite{FI94}. We have computed the spectrum of the former spin model for an
arbitrary number of particles $N$ and internal degrees of freedom $m$. Using this result and
Polychronakos's freezing trick, we have derived a simple closed-form expression for the partition
function of the FI chain. With the help of this expression, we have shown that the energy levels
can be written in terms of the standard {\em motifs} introduced by Haldane et al.~for the HS
chain~\cite{HHTBP92}, albeit with a different dispersion relation. We have performed a statistical
analysis of the chain's spectrum, showing in particular that the level density is Gaussian when
$N$ is sufficiently large and the chain's parameter $\be$ is $O(N)$. It would be desirable to find
a rigorous analytic justification of this numerical result, as has been recently done for several
spin chains of HS type whose partition function factorizes~\cite{EFG09}. We have also analyzed the
distribution of spacings between consecutive levels of the unfolded spectrum, of interest in the
context of quantum chaos. Our computations indicate that the density of spacings behaves in two
qualitatively different ways depending on the parameter $\be$. More precisely, if $\be$ is an
integer or a rational with a ``small'' denominator, the cumulative spacings distribution $P(s)$ is
approximately given by the ``square root of a logarithm'' law typical of other integrable spin
chains of HS type~\cite{BFGR08,BFGR08epl,BFGR09,BB09,BFG09}. On the other hand, for other values
of $\be$ the function $P(s)$ presents several ``cusps'', but still is neither of Poisson's nor
Wigner's type. It would be of interest in this respect to ascertain whether this behavior of
$P(s)$ is shared by the trigonometric chain of $BC_N$ type, whose spectrum also depends on a
parameter~\cite{EFGR05}.

\appendix
  \section{Existence and uniqueness of the minimum of $U$ in $C$}\label{sec.minimum}

  In this appendix we shall prove that the scalar potential $U$ in Eq.~\eqref{U} has a critical
  point in the configuration space~\eqref{C} if and only if $\be>2(N-1)$. Moreover, if this
  condition is satisfied then there is a unique critical point, which is in fact a minimum. Our
  proof is an adaptation of the argument in Ref.~\cite{CS02} for the trigonometric Sutherland
  potential.

  We shall start by expressing the logarithm of the ground state~\eqref{rho} of the scalar
  Hamiltonian~\eqref{Hsc} as
  \[
  \vp\equiv\log\rho=a\,W+W_0\,,
  \]
  where $W_0=\sum_ix_i$ and
  \[
  W=(N-\be-1)\sum_ix_i-\frac\be2\sum_i\e^{-2x_i}+\sum_{i<j}\log|\sinh(x_i-x_j)|\,.
  \]
  Since, by construction, $\Hsc\e^\vp=E_0\e^\vp$, where
  \[
  E_0=N\bigg[\ms\frac23\,a^2 (N-1) (3\be-2N+1)+2a (\be+1-N)-1\bigg]\,,
  \]
  the potential $V$ of the scalar Hamiltonian~\eqref{Hsc} can be expressed as
  \begin{equation}\label{V}
    V=(\nabla\vp)^2+\triangle\vp+E_0=\big[\nabla(a\,W+W_0)\big]^2+a\triangle W+E_0\,.
  \end{equation}
  Taking into account that
  \[
  V-a\ms\triangle W=a^2 U+2\be a\sum_i \e^{-2x_i}\,,
  \]
  equating the coefficients of $a^2$ in each side of Eq.~\eqref{V} we easily obtain
  \begin{equation}
    \label{UW}
    U=(\nabla W)^2+U_0\,,
  \end{equation}
  with
  \[
  U_0=\frac23\,N(N-1)(3 \be-2N+1)\,.
  \]
  Proceeding as in Ref.~\cite{CS02}, we next show that the Hessian of $W$ is negative definite
  everywhere. Indeed, let $\bh\equiv(h_1,\dots,h_N)\in\RR^N$. Since
  \[
  \frac{\pa^2 W}{\pa x_i\pa x_k}=-\de_{ik}\Big(2\be\e^{-2x_i}+\sum_{j;j\ne
    i}\sinh^{-2}(x_i-x_j)\Big)+(1-\de_{ik})\sinh^{-2}(x_i-x_k)\,,
  \]
  we have
  \begin{align*}
    \sum_{i,k}\frac{\pa^2 W}{\pa x_i\pa x_k}\,h_ih_k&=
    -2\be\sum_i\e^{-2x_i}h_i^2+\sum_{i\ne j}\frac{h_ih_j-h_i^2}{\sinh^{2}(x_i-x_j)}\\
    &\le -2\be\sum_i\e^{-2x_i}h_i^2+\frac12\,\sum_{i\ne j}\frac{h_j^2-h_i^2}{\sinh^{2}(x_i-x_j)}
    =-2\be\sum_i\e^{-2x_i}h_i^2\,,
  \end{align*}
  which is clearly negative definite in $\bh$ for all $\bx\in C$. Since, by Eq.~\eqref{UW},
  \begin{equation}
    \label{Ui}
    \frac{\pa U}{\pa x_i}=2\sum_k\frac{\pa^2 W}{\pa x_i\pa x_k}\,\frac{\pa W}{\pa x_k}\,,
  \end{equation}
  it follows that $\bxi\in C$ is a critical point of $U$ if and only if it is a critical point of
  $W$, i.e., if and only if
  \[
  \sum_{j;j\ne i}\coth(\xi_i-\xi_j)=\be-N+1-\be\e^{-2\xi_i}\,,\qquad i=1,\dots,N\,.
  \]
  Setting
  \begin{equation}\label{zexi}
    \ze_{N-i+1}=\be\e^{-2\xi_i}\,,\qquad 0<\ze_1<\cdots<\ze_N\,,
  \end{equation}
  we can easily rewrite the previous system as
  \begin{equation}
    \label{zesys}
    \sum_{j;j\ne i}\frac{2\ze_i}{\ze_j-\ze_i}=\ze_i-\be+2(N-1)\,,\qquad i=1,\dots,N\,.
  \end{equation}
  From the last of these equations it follows that
  \[
  \be=2(N-1)+\ze_N+\sum_{j=1}^{N-1}\frac{2\ze_N}{\ze_N-\ze_j}>2(N-1)\,,
  \]
  so that~Eq.~\eqref{cond} is a necessary condition for $W$, and hence $U$, to have a critical
  point in $C$.

  Conversely, if Eq.~\eqref{cond} is satisfied then $W$ must have a unique critical point (a
  maximum) in $C$. Indeed, since
  \begin{align*}
    W &= (N-\be-1)\sum_i
    x_i-\frac\be2\,\sum_i\e^{-2x_i}+\sum_{i<j}(x_i+x_j)+\sum_{i<j}\log\big|\e^{-2x_i}-\e^{-2x_j}\big|\\
    &=(2N-\be-2)\sum_ix_i-\frac\be2\,\sum_i\e^{-2x_i}+\sum_{i<j}\log\big|\e^{-2x_i}-\e^{-2x_j}\big|\,,
  \end{align*}
  it follows that when $\be>2(N-1)$ the prepotential $W$ tends to $-\infty$ both on the boundary
  of the set $C$ and for $x_i\to\pm\infty$. Hence $W$ must have \emph{at least} a maximum in $C$. On
  the other hand, all the critical points of $W$ in $C$ must be maxima, since we have just seen
  that the Hessian of $W$ is negative definite everywhere. Hence $W$ has \emph{at most} a relative
  maximum in $C$, thus establishing our claim.

  The previous result implies that if $\be>2(N-1)$ then $U$ has a unique critical point $\bxi$ in
  $C$. To ascertain its nature, it suffices to note that from Eq.~\eqref{Ui} we easily have
  \[
  \frac{\pa^2 U}{\pa x_i\pa x_j}(\bxi)=2\sum_k\frac{\pa^2 W}{\pa x_i\pa x_k}(\bxi)\, \frac{\pa^2
    W}{\pa x_k\pa x_j}(\bxi)\,,
  \]
  so that the Hessian of $U$ at $\bxi$ is the square of that of $W$. Since the latter Hessian is
  negative definite, this shows that the Hessian of $U$ at $\bxi$ is positive definite, so that
  $\bxi$ is a minimum.

  Note, finally, that the zeros of the generalized Laguerre polynomial $L_N^{\be-2N+1}$ are known
  to satisfy Eq.~\eqref{zesys}, cf.~Ref.~\cite{Sz75}. {}From this fact and Eq.~\eqref{zexi} it
  follows that the coordinates $\xi_i$ of the unique minimum of $U$ in $C$ are given by
  Eq.~\eqref{zei}.

\section{Computation of the mean and variance of the chain's energy}\label{sec.musigma}

In this appendix we shall compute in closed form the mean $\mu$ and the variance $\si^2$ of
the spectrum of the spin chain~\eqref{cH} in terms of the number of particles $N$ and
internal degrees of freedom $m$. We shall start with the mean energy $\mu=m^{-N}\tr\cH$. Using the
formulas for the traces of the spin permutation operators $S_{ij}$ in Ref.~\cite{EFGR05} and
Eq.~\eqref{cEmaxF} we easily obtain
\begin{equation}\label{mu}
  \mu=\Big(1-\frac\vep m\Big)\sum_{i\neq j} h_{ij}=\frac1{12}\,\Big(1-\frac\vep m\Big)N(N-1)(3\be-4N+2)\,.
\end{equation}
Consider next the variance of the energy
\[
\si^2=m^{-N}\tr(\cH^2)-\mu^2\,,
\]
which is independent of $\vep$ on account of the identity~\eqref{cHs}. Proceeding as in
Ref.~\cite{FG05} we readily obtain
\begin{equation}\label{si2def}
\si^2=2\Big(1-\frac1{m^2}\Big)\sum_{i\ne j}h_{ij}^2\,.
\end{equation}
The last sum can be evaluated using the procedure of Ref.~\cite{BFGR08}, as we shall now show.
Note, to begin with, that
\begin{equation}\label{hij2}
  \sum_{i\ne j}h_{ij}^2=\frac12\sum_{i\ne
    j}\frac{\ze_i^4+\ze_j^4-(\ze_i^2-\ze_j^2)^2}{(\ze_i-\ze_j)^4}
  = \sum_{i\ne
    j}\frac{\ze_i^4}{(\ze_i-\ze_j)^4}-\sum_{i\ne j}\frac{\ze_i^2}{(\ze_i-\ze_j)^2}-
  \sum_{i\neq j} h_{ij}\,.
\end{equation}
In order to evaluate the second of these sums, we use the following identity from
Ref.~\cite{Ah78}:
\[
  \sum_{j,j\neq i}\frac{12\,\ze_i^2}{(\ze_i-\ze_j)^2}
  =-\ze_i^2+2(\be+2)\ze_i-(\be-2N+2)(\be-2N+6)\,.
\]
Summing over $i$ and using the identities
\begin{equation}\label{zeze2}
\sum_i\ze_i =N(\be-N+1)\,,\qquad \sum_i\ze_i^2 =N\be(\be-N+1)
\end{equation}
(cf.~Ref.~\cite{BFGR08}) we easily obtain
\begin{equation}\label{sum2}
\sum_{i\ne j}\frac{\ze_i^2}{(\ze_i-\ze_j)^2}=\frac{N}{12}\,(N-1)(3\be-4N+8)\,.
\end{equation}
Consider next the first sum in the RHS of Eq.~\eqref{hij2}. {}From Theorem 5.1 of Ref.~\cite{AM83},
after a long but straightforward calculation we obtain
\begin{align}
  \sum_{j\neq
    i}\frac{720\,\ze_i^4}{(\ze_i-\ze_j)^4}=&\sum_i\ze_i^4-4(\be+2)\sum_i\ze_i^3\notag\\
   &+\big[8N^2-8N(\be+1)+2(3\be+4)(\be+2)\big]\sum_i\ze_i^2\notag\\
   &-4(\be+2)\big[4N^2-4N(\be+1)+\be(\be+2)-18\big]\sum_i\ze_i\notag\\
   &-N(\be-2N+2)\big[8N^3-4N^2(3\be+2)+N(6\be^2+8\be-216)\notag\\
   &\phantom{-N(\be-2N+2)\big[8}-\be(\be^2+2\be-108)+360\big]\label{sum4}\,.
\end{align}
All the sums in the RHS of this equation can be evaluated from Eq.~\eqref{zesys}.
Indeed, multiplying this equation by $\ze_i^2$ and summing over $i$ we have
\begin{align*}
  \sum_{i\neq j}\frac{2\ze_i^3}{\ze_i-\ze_j}&=\sum_{i\neq
    j}\frac{\ze_i^3-\ze_j^3}{\ze_i-\ze_j}=\sum_{i\neq j}(\ze_i^2+\ze_i\ze_j+\ze_j^2)\\&
  =(2N-3)\sum_i\ze_i^2+\Big(\sum_i\ze_i\Big)^2=\sum_i\ze_i^3-(\be-2N+2)\sum_i\ze_i^2\,,
\end{align*}
and hence, by the identities~\eqref{zeze2},
\begin{equation}\label{ze3}
  \sum_i\ze_i^3=N \big[N^3-2 N^2 (\be+1 )+N (3 \be+1 )+\be(\be^2-1)\big]\,.
\end{equation}
Similarly, multiplying~\eqref{zesys} by $\ze_i^3$ and summing over $i$ we obtain
\begin{align*}
  \sum_{i\neq j}\frac{2\ze_i^4}{\ze_i-\ze_j}&=\sum_{i\neq
    j}\frac{\ze_i^4-\ze_j^4}{\ze_i-\ze_j}=\sum_{i\neq
    j}(\ze_i^3+\ze_i^2\ze_j+\ze_i\ze_j^2+\ze_j^3)\\&
  =2(N-2)\sum_i\ze_i^3+2\Big(\sum_i\ze_i^2\Big)\Big(\sum_i\ze_i\Big)
  =\sum_i\ze_i^4-(\be-2N+2)\sum_i\ze_i^3\,,
\end{align*}
so that, by Eqs.~\eqref{zeze2} and~\eqref{ze3},
\begin{equation}\label{ze4}
  \sum_i\ze_i^4=N (N-\be-1)\big[N^2 (3 \be-2 )-N(\be+1)(3\be-2)-\be(\be-1)(\be-2)\big]\,.
\end{equation}
Substituting Eqs.~\eqref{ze3} and~\eqref{ze4} into Eq.~\eqref{sum4} we easily arrive at
\begin{multline}
  \label{sum4final}
  \sum_{j\neq i}\frac{\ze_i^4}{(\ze_i-\ze_j)^4}= \frac{N}{720}\,(N-1) \big[16 N^3+N^2 (6-25
  \be)\\+N (10 \be^2-35 \be-454)+25 \be^2+350 \be+576\big]\,.
\end{multline}
The previous identity, together with Eqs.~\eqref{si2def}, \eqref{hij2}, and \eqref{sum2}, finally
yield the explicit formula~\eqref{si2final} for the variance of the energy of the FI chain.

\medskip

\begin{acknowledgments}
  This work was partially supported by the Spanish Ministry of Science and Innovation under grant
  No.~FIS2008-00209, and by the Universidad Complutense and Banco Santander under grant
  No.~GR58/08-910556. J.C.B.~acknowledges the financial support of the Spanish Ministry of Science
  and Innovation through an FPU scholarship.\vspace*{.5cm}
\end{acknowledgments}

\end{document}